\documentclass[useAMS,usenatbib]{mn2e}
\usepackage{amssymb,latexsym,amscd,amsmath,graphicx,ifpdf,url}
\usepackage{longtable}

\def\be{\begin{equation}}
\def\ee{\end{equation}}
\def\ba{\begin{array}}
\def\ea{\end{array}}
\def\bea{\begin{eqnarray}}
\def\eea{\end{eqnarray}}
\def\bi{\begin{itemize}}
\def\ei{\end{itemize}}

\def\half{{\textstyle{1\over2}}}

\newcommand{\prc}{{\it Physical Review C}}

\newcommand{\apj}{{\it The Astrophysical Journal }}

\newcommand{\apjl}{{\it The Astrophysical Journal Letters }}

\newcommand{\epja}{{\it European Physical Journal A}}
\newcommand{\mnras}{{\it Monthly Notices of the Royal Astronomical Society }}

\title{Efficacy of crustal superfluid neutrons in pulsar glitch models}
\author[J. Hooker, W. G. Newton and Bao-An Li]{J. Hooker, W. G. Newton and Bao-An Li\\
Department of Physics and Astronomy, Texas A\& M University - Commerce, Commerce, Texas, 75429-3011}
\begin{document}

\date{\today}

\pagerange{\pageref{000}--\pageref{000}} \pubyear{0000}

\maketitle

\label{firstpage}

\begin{abstract}

In order to assess the ability of purely crust-driven glitch models to match the observed glitch activity in the Vela pulsar, we conduct a systematic analysis of the dependence of the fractional moment of inertia of the inner crustal neutrons on the stiffness of the nuclear symmetry energy at saturation density $L$. We take into account both crustal entrainment and the fact that only a fraction $Y_{\rm g}$ of the core neutrons may couple to the crust on the glitch-rise timescale. We use a set of consistently-generated crust and core compositions and equations-of-state which are fit to results of low-density pure neutron matter calculations. When entrainment is included at the level suggested by recent microscopic calculations and the core is fully coupled to the crust, the model is only able to account for the Vela glitch activity for a 1.4$M_{\odot}$ star if the equation of state is particularly stiff $L>100$ MeV. However, an uncertainty of about 10\% in the crust-core transition density and pressure allows for the Vela glitch activity to be marginally accounted for in the range $L\approx30-60$MeV consistent with a range of experimental results. Alternatively, only a small amount of core neutrons need be involved. If less than 50\% of the core neutrons are coupled to the crust during the glitch, we can also account for the Vela glitch activity using crustal neutrons alone for EOSs consistent with the inferred range of $L$. We also explore the possibility of Vela being a high-mass neutron star, and of crustal entrainment being reduced or enhanced relative to its currently predicted values.

\end{abstract}

\begin{keywords}
dense matter - equation of state - stars:neutron - (stars:) pulsars: general - (stars:) pulsars: individual: Vela
\end{keywords}

\section{Introduction}

The rotational evolution of young pulsars ($\lesssim10^7$ years old) is often observed to be interrupted by glitches - sudden increases in spin frequency $\nu$. The range of glitch sizes $\frac{\Delta \nu}{\nu_{0}} \sim 10^{-11}$ to $10^{-5}$ and modes of post-glitch recovery is observed to be quite diverse across the pulsar population \citep{Espinoza:2011pq}. One of the most studied pulsars, Vela, quasi-periodically produces large glitches $\frac{\Delta \nu}{\nu_{0}} \sim 10^{-6}$ \citep{Melatos:2007gn}; the relatively large number of glitches observed from Vela (21) has made it a test-bed for proposed glitch mechanisms. Much effort has been devoted to examining two-component glitch models, in which one component of the neutron star is for most of the time decoupled from the solid crustal lattice to which the magnetic field lines are anchored and whose rotational evolution we directly observe. This decoupled component acts as an angular momentum reservoir, and occasionally re-couples to the rest of the star, transferring some of its angular momentum and spinning the star up \citep{Anderson:1975zze,Alpar:1977aa}. The most studied class of two-component models posit that the angular momentum reservoir consists of the free superfluid neutrons that permeate the inner crust. The mechanism by which they are decoupled from the core, and periodically recoupled, is still a matter of considerable uncertainty owing to the difficulty in marrying the complex microphysics with the global rotational dynamics of the fluid components. In the pinning paradigm, the neutron superfluid vortices in the crust could interact with the lattice of nuclei to become `pinned' to the crust \citep{Pizzochero:1997iq,Avogadro:2008uy}, therefore preventing them from moving radially outwards from the rotation axis in response to the secular spin-down of the star \citep{Alpar:1977aa,Pines:1980a,Anderson:1982a,Alpar:1984b}. They thus become decoupled from the rotational evolution of the charged component of the crust and that part of the core that couples strongly to it. As a lag builds up between the angular frequency of the charged component (that which we observe) and the frequency of the crustal superfluid, the Magnus force acting radially outwards on the vortices grows, until eventually it overcomes the pinning force, and the vortices unpin \emph{en masse} and couple to the charged component, transferring angular momentum. Despite recent advances in combining the microphysics and hydrodynamics of this model \citep{Sidery:2009at,vanEysden:2010ha,Haskell:2011xe,Pizzochero:2011dd,Seveso:2012ts} certain aspects of the model such as the exact unpinning trigger and a convincing account of the process by which a large number of vortices subsequently moves remain to be addressed \citep{Link:1996a,Glampedakis:2008mi,Warszawski:2008fb, Melatos:2009au, Warszawski:2012wa, Warszawski:2012ns}. In addition, it is possible that pinning of neutron vortices occurs elsewhere in the star \citep{Srinivasan1990a,Jones1991a,Mendell1991a,Ruderman:1998a,Link:2003hq,Link2012a}, or that at least some glitches arise by other means such as hydrodynamical instabilities and turbulence \citep{Melatos2007a,Andersson2004a}.

In this paper we consider the class of models in which the free inner crustal neutrons are the component which acts as the effective angular momentum reservoir. On can analyze the tenability of these models in general terms relatively free of the uncertainties outlined above. The observed glitch activity of the Vela pulsar can be used to infer the fractional moment of inertia of the angular momentum reservoir $\Delta I$ compared to that of the portion of the star it couples to at the time of glitch, $I$ (which we shall referred to as the charged component of the star), $\Delta I/I \gtrsim 1.6 \%$ \citep{Link:1999ca,Espinoza:2011pq}. This is the most stringent constraint from any pulsar, with the next largest glitch activity observed coming from PSR J0537-6910 $\Delta I/I \gtrsim 0.9 \%$ \citep{Espinoza:2011pq,Andersson:2012iu} from which we have observed the most glitches of any pulsar.

Assuming an upper limit to the moment of inertia of the angular momentum reservoir to be that of the entire crustal neutron superfluid $\Delta I = I_{\rm csf}$, and that the charged component it couples to is essentially the whole star $I = I_{\rm tot}$ (i.e. the entire core couples to the charged component of the crust on glitch-rise timescales), many realistic neutron star (NS) equations of state (EoSs) can satisfy the condition $\Delta I/I \gtrsim 1.6 \%$, and constraints can be placed on the NS EoS \citep{Lorenz:1992zz,Link:1999ca,Fattoyev:2010tb}. However, recent calculations of the strength of the entrainment of superfluid neutrons by the crustal lattice via Bragg scattering suggest that only a fraction of the crustal neutrons are effectively free, and consequently the upper limit to the moment of inertia of the angular momentum reservoir is reduced: $\Delta I \sim 0.2 I_{\rm csf}$. Initial analyses suggested that makes $\Delta I /I$ too small to explain the observed glitch sizes \citep{Chamel:2012ae,Chamel:2012zn,Andersson:2012iu}, and suggests one must at least involve other components of the star in addition to the crust superfluid as the store of angular momentum. These studies assume a strong coupling between crust and core so that $I \sim I_{\rm tot}$. More recently, however, analyses using a wider range of possible equations-of-state (EOSs) suggest that a crustal angular momentum reservoir might still be consistent with Vela glitch observations \citep{Piekarewicz2014a,Steiner2015a}.

However, estimates of the crust-core coupling timescales due to interactions of neutron vortices with core protons (which may form a Type I or Type II superconductor) \citep{Alpar:1988a,Sedrakian:2004yq,Andersson:2005rs,Jones:2006a,Babaev:2009a} raise the possibility of only a small fraction of the core neutrons being coupled to the crust on the glitch rise timescale $<40$s \citep{Dodson:2002gy}. Therefore it is possible that $I \ll I_{\rm tot}$, allowing the ratio $\Delta I /I$ to satisfy the lower bound of 1.6\% again with only crustal superfluid neutrons involved for a wider range of EOSs.

The aim of this paper is to examine the range of predictions for $\Delta I/I$ taking into account systematically variations in the most uncertain nuclear matter parameters over their experimentally and theoretically constrained ranges, and crustal entrainment in a way that is consistent with the crust models using the recently calculated values in \citep{Chamel:2012zn}. We will also go beyond similar studies by allowing for a fraction $Y_{\rm g}<1$ of core neutrons to be coupled to the crust at the time of glitch. We shall apply systematically and consistently generated sequences of crust and core EOSs together with the relevant crust compositions \citep{Newton:2011dw} to calculate the relevant moments of inertia including that of the free crustal neutrons explicitly. The consistent modeling of crust and core properties when exploring the dependence of neutron star observables has been presented before \citep{Gearheart:2011qt,Wen:2011xz}, and here extends to modeling the crust thickness, density of superfluid neutrons throughout the crust, core EOS and core proton fraction using the same underlying nuclear matter EOSs. 

Much effort has been devoted to constraining the EOS of nuclear and neutron star matter, particularly through constraining the density dependence of the symmetry energy at nuclear saturation density $n_0$, parameterized by $L = 3n_0 p_0$ where $p_0$ is the pressure of pure neutron matter at saturation density, which is strongly correlated with the pressure in neutron stars at that density. Nuclear experimental probes (e.g. \cite{Li:2008gp,Tsang:2012se}) give a conservative range $L=20-120$ MeV, although some more recent results on the nuclear experimental side \citep{Lattimer:2012xj}, as well as tentative constraints from neutron star observation \citep{Newton:2009vz,Gearheart:2011qt,Wen:2011xz,Steiner:2011ft} and from \emph{ab-initio} pure neutron matter calculations \citep{Gezerlis:2009iw, Hebeler:2009iv, Gandolfi:2011xu} tend to favor the lower half of that range (although, for a counter-example, see e.g. \cite{Sotani:2012qc}). Particularly, combining inferred values of $L$ from several nuclear experimental probes suggests a range $L\approx30-60$ MeV \citep{Hebeler2013a,Lattimer2014a}. The high-density behavior of the EOS is even more uncertain both theoretically and experimentally \citep{Xiao:2009zza,Russotto:2011hq}, even if one restricts the composition to purely nucleonic matter, with some of the only constraints coming from analysis of heavy-ion collisions \citep{Danielewicz:2002pu} and the recent observations of  $\sim 2 M_{\odot}$ neutron stars \citep{Demorest:2010bx,Antoniadis:2013}. In this paper we shall explore the impact of systematically varying the density dependence of the symmetry energy $L$ at saturation density on the the predictions for $I_{\rm csf}/I$. 

In Section 2 we describe our glitch modeling and set of EOSs and our calculation of observable quantities. In Section 3 we present and discuss our results and in Section 4 we discuss our conclusions.

%===========================================================>
%
%					FIGURE 1
%
%===========================================================>
\begin{figure*}\label{fig:1}
\begin{center}
\includegraphics[width=8.5cm,height=5cm]{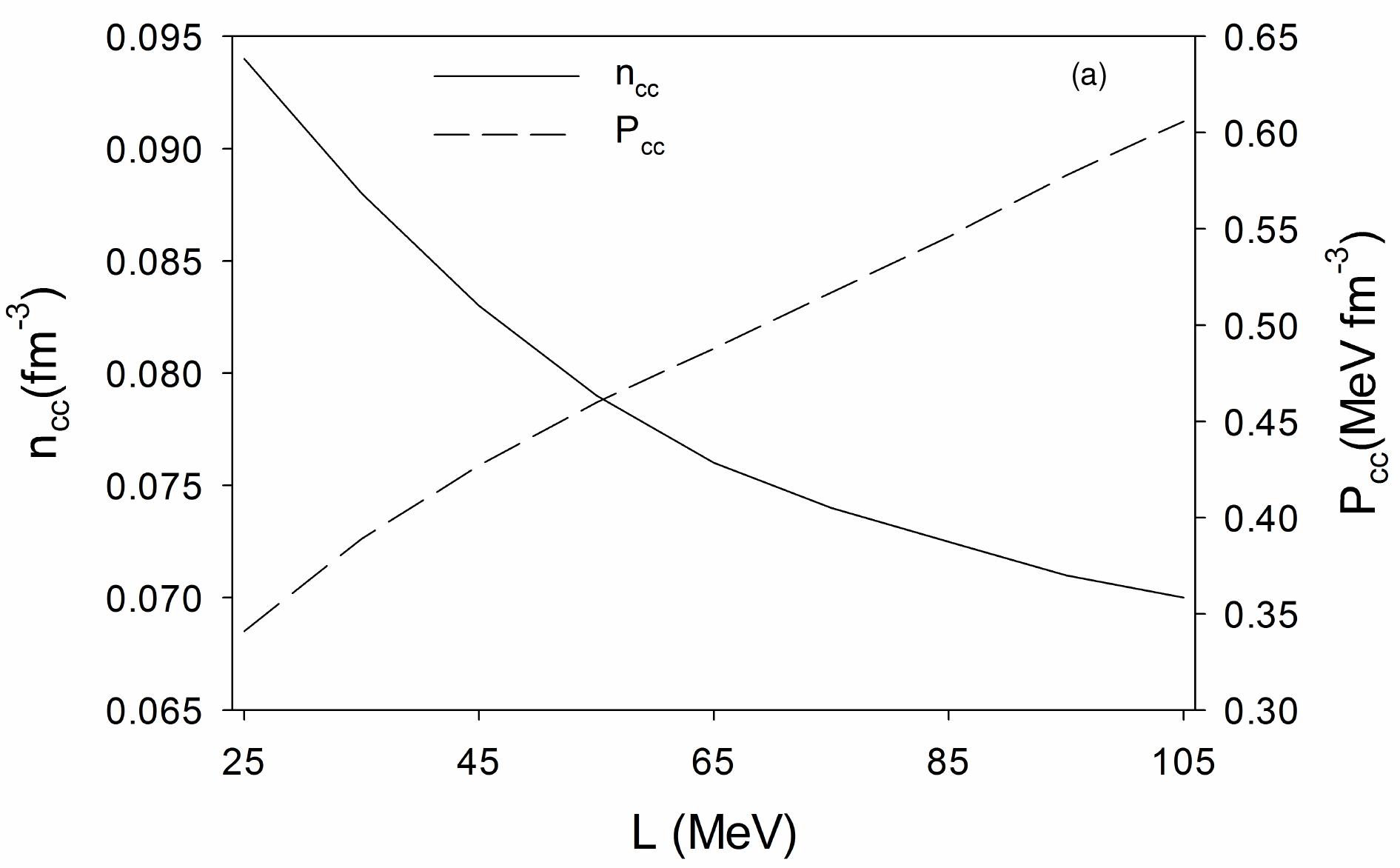}\hspace{0.5cm}\includegraphics[width=8cm,height=5cm]{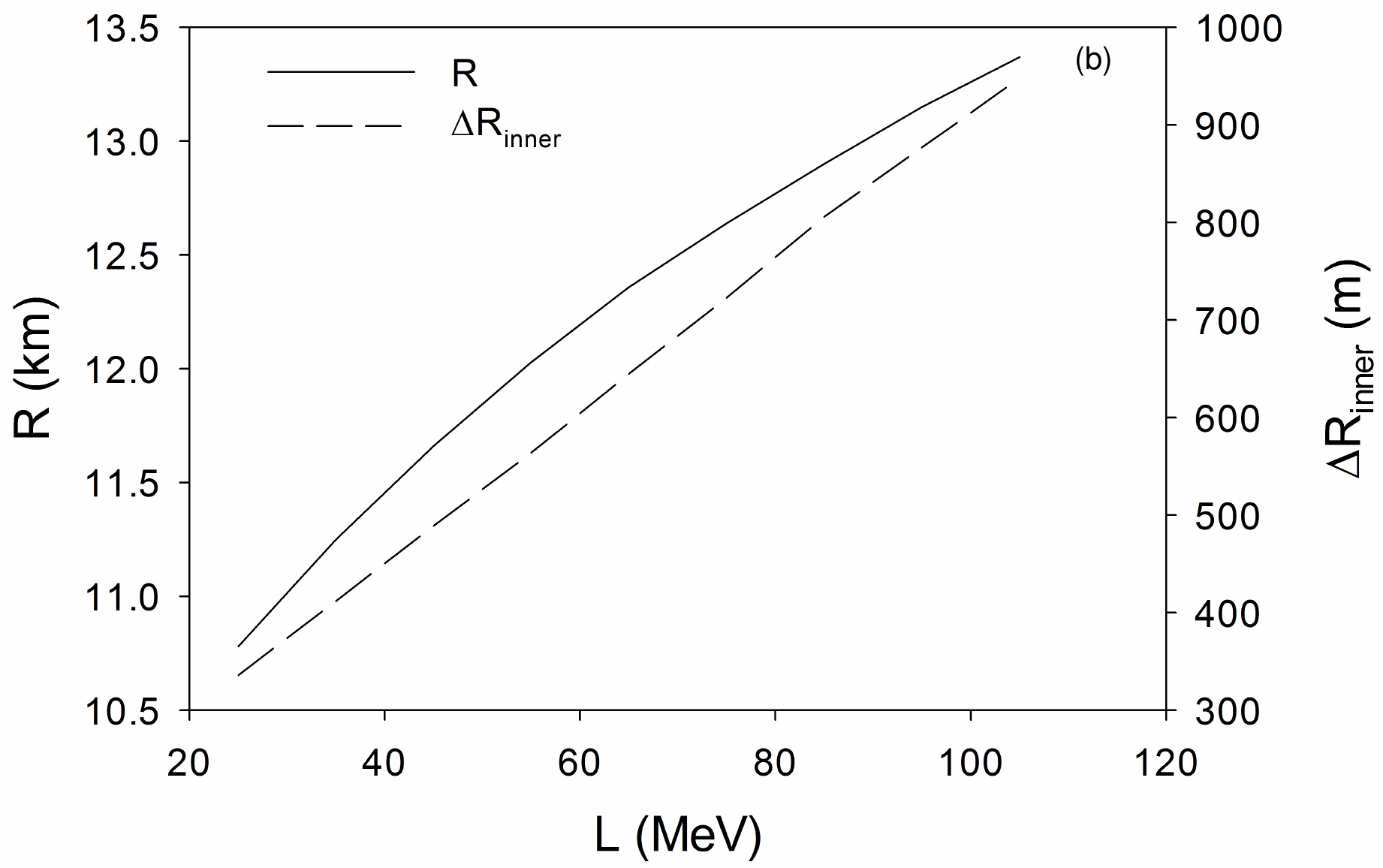}
\caption{Crust-core transition density $n_{\rm cc}$ and pressure $P_{\rm cc}$ (a), and the radius $R$ and inner crust thickness $\Delta R_{\rm inner}$ (b) of a 1.4$M_{\odot}$ neutron star as a function of the density-slope of the symmetry energy $L$ for the sequence of EOSs used in our study.}
\end{center}
\end{figure*}
%===========================================================>

\section{The Glitch Model}

The observed angular frequency, $\Omega$, of a pulsar is presumed to be that of its ionic crustal lattice in which the magnetic field lines are anchored. When considering glitch sizes and immediate post-glitch evolution, it is important to define that component of the star strongly coupled to the lattice on timescales comparable with the glitch rise time, which is observationally constrained to be $\lesssim$ 40s \citep{Dodson:2002gy}. In our minimal model of the core which contains purely nucleonic matter, this component contains the core protons and some fraction of the core neutrons, and we shall refer to it as the charged component of the star.

The moment of inertia of a star of radius $R$ in the limit of small angular frequency $\Omega$ \citep{Hartle:1968si} is given by

\begin{equation} \label{eq:MoI1}
I_{\rm tot}=\frac{8 \pi}{3} \int_0^R r^{4}e^{-\nu(r)}\frac{\bar{\omega}(r)}{\Omega}\frac{\left( \mbox{\Large{$ \varepsilon $}} (r)+P(r) \right)}{\sqrt{1-2GM(r)/r}}\mathrm{d}r,
\end{equation}

\noindent where $\text{\Large{$ \varepsilon $}}(r)$ is energy density of matter in the star, $P(r)$ is the pressure and $M(r)$ is the mass contained in radius $r$. $\nu(r)$ is a radially-dependent metric function given by

\begin{equation} \label{eq:MoI2}
\nu(r)=\frac{1}{2} \ln \left( 1- \frac{2GM}{R} \right) - G \int_r^R \frac {\left( M(x)+4 \pi x^{3} P(x) \right)}{x^{2} \left( 1 - 2 GM(x)/x \right)} \mathrm{d}x,
\end{equation}

\noindent and $\bar{\omega}$ is the frame dragging angular velocity

\begin{equation} \label{eq:MoI3}
\frac{1}{r^{3}}\frac{\mathrm{d}}{\mathrm{d}r} \left( r^{4}j(r)\frac{\mathrm{d}\bar{\omega}(r)}{\mathrm{d}r} \right) + 4 \frac{\mathrm{d}j(r)}{\mathrm{d}r} \bar{\omega}(r)=0,
\end{equation}

\noindent where

\begin{equation} \label{eq:fp3}
j(r)=e^{-\nu(r)-\lambda(r)}=\sqrt{1-2GM(r)/r} e^{-\nu(r)}
\end{equation}

\noindent for $r \leq R$. 

The charged component of the star includes the crust lattice plus the protons in the core and some fraction of core superfluid neutrons. Neutron vortices can entrain protons and become magnetized, allowing electrons to scatter off them and therefore coupling to the charged component on timescales of $\sim 10 - 1000$s for the Vela pulsar \citep{Alpar:1984b,Alpar:1988a}. In regions where protons are Type II superconductors, they form fluxtubes to which neutron vortices can become pinned, coupling them to the charged component on timescales of days or longer \citep{Sedrakian:1995a,Sedrakian:2004yq,Babaev:2009a}. By comparing the above estimates of coupling timescales with the upper limit of the glitch rise time $\lesssim 40$s, one may infer that only some fraction of the core neutron superfluid will contribute to the charged component of the star at the time of glitch. That fraction is quite uncertain, and enters into the model as a free parameter $Y_{\rm g}$, but above estimates indicate that it is possible to have $Y_{\rm g} \ll 1$ \citep{Haskell:2011xe}. We also denote the total neutron fraction of the core at a given radius $r$ by $Q(r)$, determined by the EOS. Then the moment of inertia of the charged component can be expressed \citep{Seveso:2012ts}

%===========================================================>
%
%					FIGURE 2
%
%===========================================================>
\begin{figure*}\label{fig:2}
\begin{center}
\includegraphics[width=6cm,height=5cm]{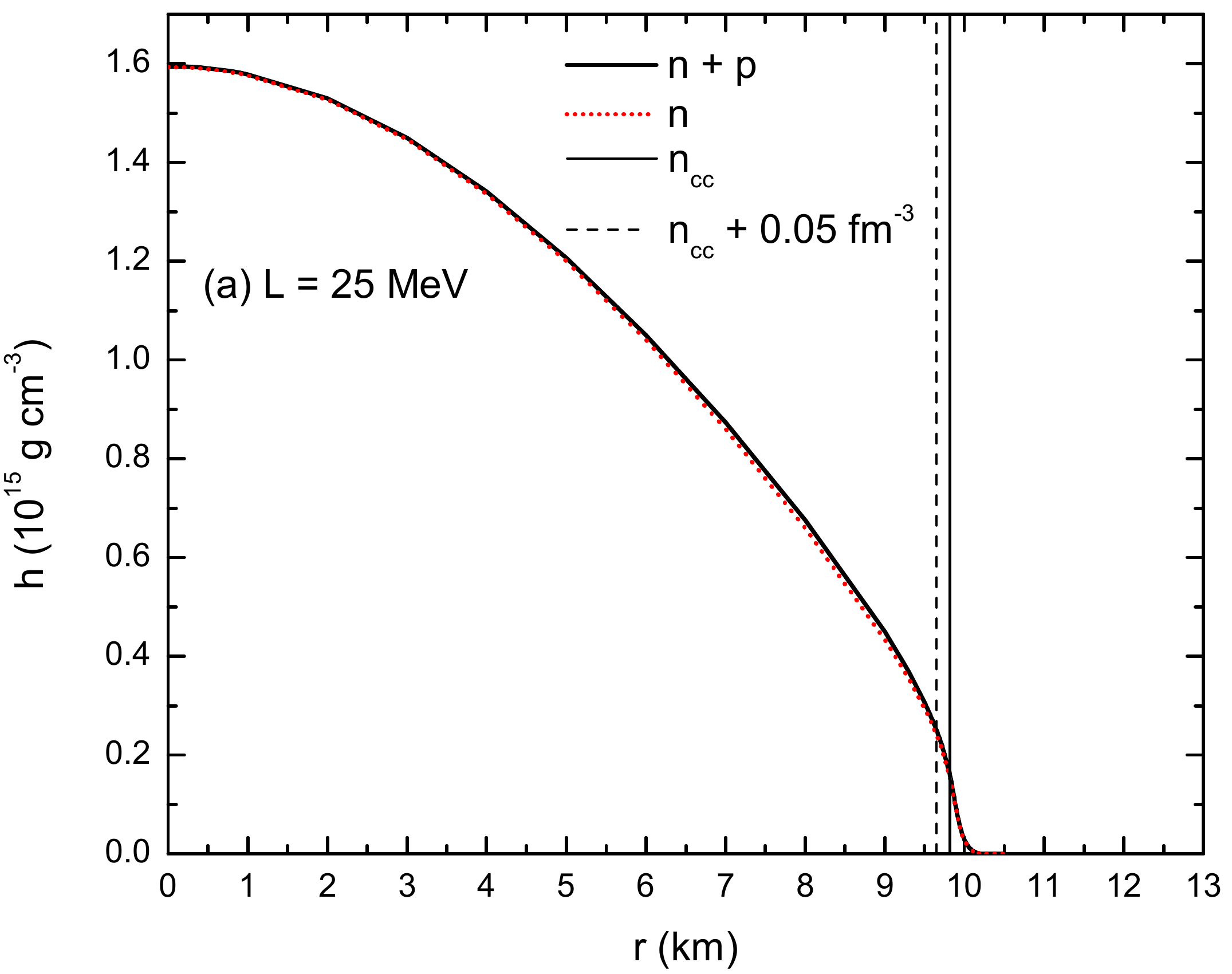}\includegraphics[width=6cm,height=5cm]{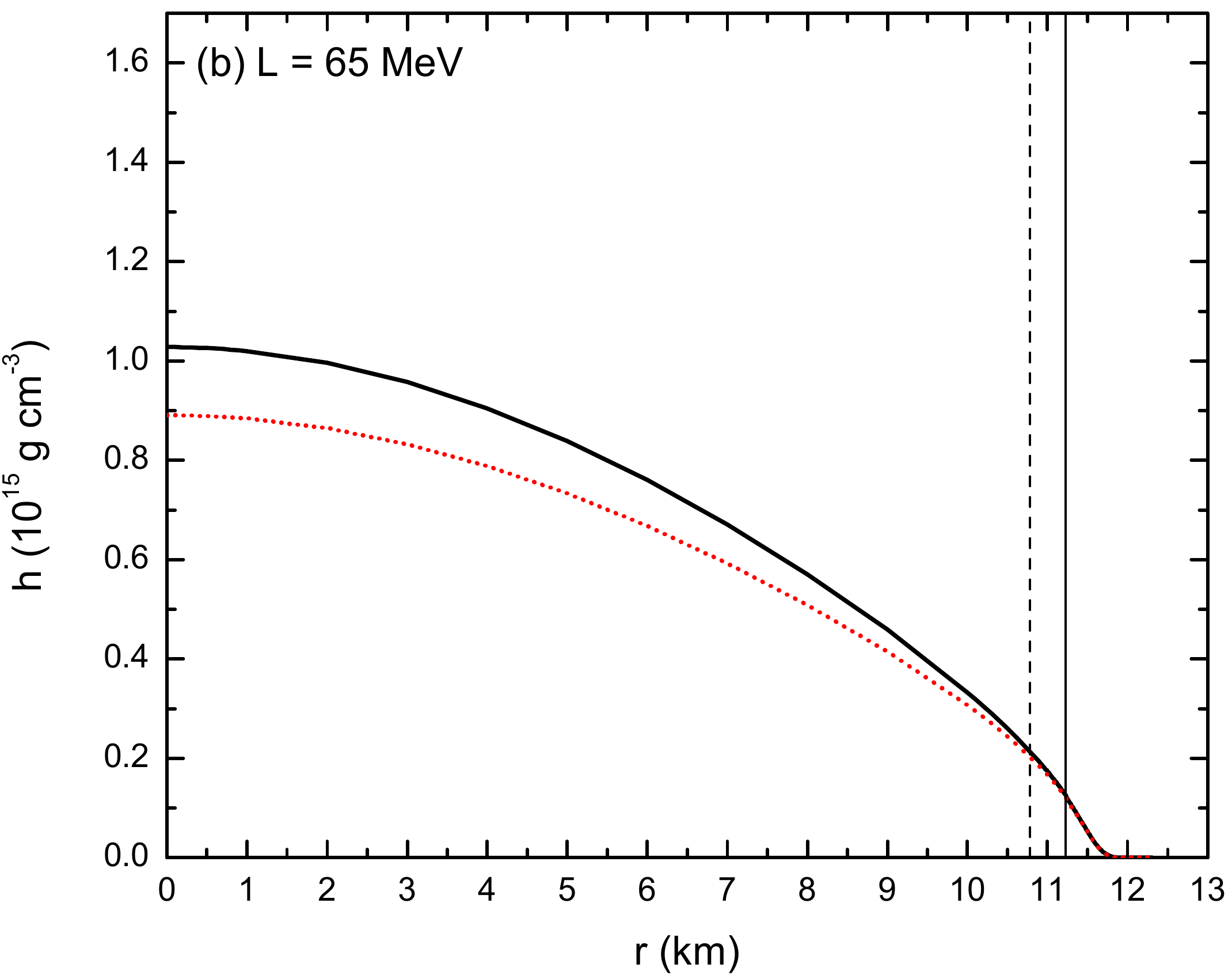}\includegraphics[width=6cm,height=5cm]{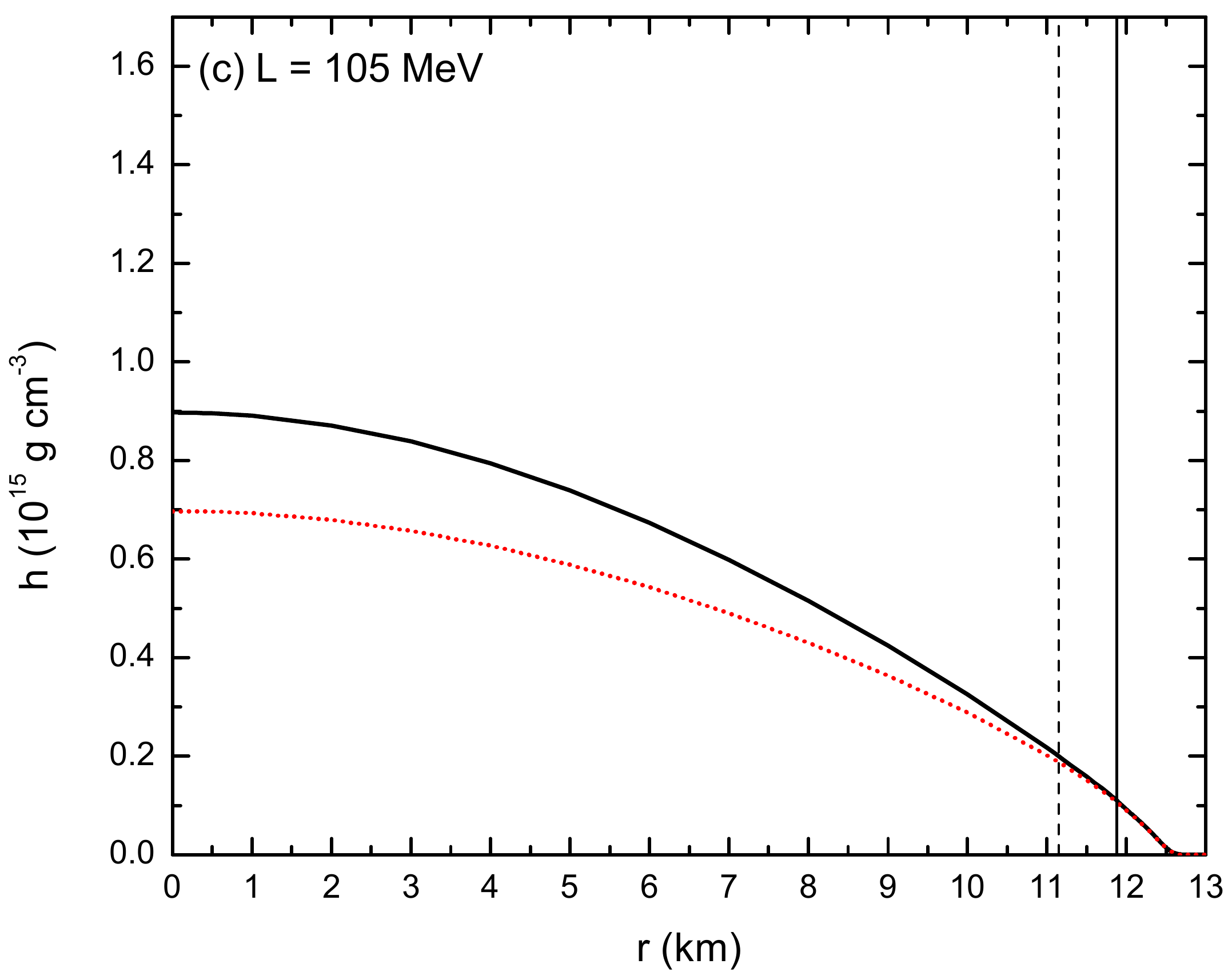}
\caption{The total enthalpy density (black bold line) and that of the free neutrons (dotted red line) as a function of radial distance from the center of a $1.4 M_{\odot}$ neutron star for values of the saturation density symmetry energy stiffness $L=25$ MeV (a), $L=65$ (b), and $L=105$ MeV (c). The radial distance corresponding to the crust-core transition density is shown by the vertical thin solid line, and that corresponding to a density of 0.05 fm$^{-3}$ above the transition density is shown by the vertical dashed line.}
\end{center}
\end{figure*}
%===========================================================>

\begin{equation} \label{eq:MoI1}
I_{\rm c}=\frac{8 \pi}{3} \int_0^R r^{4} [1 - Q(r) (1 - Y_{\rm gl})] e^{-\nu(r)}\frac{\bar{\omega}(r)}{\Omega}\frac{\left( \mbox{\Large{$ \varepsilon $}} (r)+P(r) \right)}{\sqrt{1-2GM(r)/r}}\mathrm{d}r.
\end{equation}

The total moment of inertia of superfluid neutrons in the inner crust of the star is given by

\begin{equation} \label{eq:new1}
I_{\rm csf}=\frac{8 \pi}{3} \int_{R_{\rm inner}}^{R_{\rm outer}} r^{4}e^{-\nu(r)}\frac{\bar{\omega}(r)}{\Omega}\frac{\left( \mbox{\Large{$ \varepsilon $}}_{n} (r)+P_{n}(r) \right)}{\sqrt{1-2GM(r)/r}}\mathrm{d}r
\end{equation}

\noindent where $\mbox{\Large{$ \varepsilon $}}_{n} (r)$ is the energy density of crustal superfluid neutrons, $P_{n}(r)$ is the pressure of the crustal superfluid neutrons and $R_{\rm inner}$ and $R_{\rm outer}$ are the radius boundaries for the inner crust.

Entrainment of superfluid neutrons by the crust's lattice reduces the mobility of the neutrons with respect to that lattice. It can be shown that this effect is encoded by introducing an effective ``mesoscopic'' neutron mass $m_{\rm n}^*$ \citep{Chamel:2004in,Chamel:2005my,Chamel:2012zn}; larger values correspond to stronger coupling between the neutron superfluid and the crust, and a reduction in the fraction of superfluid neutrons able to store angular momentum for the glitch event. One can include this effect by modifying the integrand Eq.~\ref{eq:new1} by multiplying by a factor of 
${m_{\rm n} / m_{\rm n}^* (r)}$ where $m_{\rm n}^*(r)$ is the effective mass at radius $r$ in the crust. 

We obtain $m_{\rm n}^*(r)$ from the results of Chamel \citep{Chamel:2012zn} by interpolating between the values calculated at specific densities to find the effective mass at arbitrary locations in the inner crust.

It is yet to be determined how entrainment is modified when one includes hitherto neglected effects such as the spin-orbit interaction in the calculations\citep{Chamel:2012zn}. In order to account for possible variations in the strength of entrainment, and to systematically trace the effect of introducing entrainment, we introduce a parameter $e$ which we use to control the strength of the entrainment: 

\be
{m_{\rm n}^* \over m_{\rm n}} \to 1 + \left({m_{\rm n}^* \over m_{\rm n}} - 1\right)e
\ee

\noindent where $e=0$ corresponds to no entrainment and $e = 1$ corresponds to the full strength entrainment as predicted by \citep{Chamel:2012zn}. We will consider also entrainment reduced by 50\% $e$=0.5 and entrainment enhanced by 50\% $e$=1.5.

The analysis of \citep{Link:1999ca} of the Vela pulsar glitches, recently updated to include the most recently observed glitches \citep{Espinoza:2011pq} identifies the minimum amount of angular momentum stored in the crustal superfluid reservoir $I_{\rm csf}$ relative to that of the charged component of the star $I_{\rm c}$ with the parameter $G$, defined by

\be
G \equiv \frac{I_{\rm csf}}{I_{\rm c}} \geq \frac{\bar{\Omega}}{|\dot{\Omega}|} \mathcal{A} = 0.016
\ee

\noindent where $\mathcal{A}$ is the glitch activity parameter of the pulsar, the fraction of frequency spin-down reversed by glitches determined by a linear fit to a plot of the cumulative relative glitch size over time \citep{Link:1999ca}. 

\subsection{Nuclear matter parameters and crust and core equations of state}

The microphysical ingredients in the glitch model include the total pressure and energy density $P(n_{\rm b})$, $\mbox{\Large{$ \varepsilon $}}(n_{\rm b})$ and those of the superfluid neutrons $P_{\rm n} (n_{\rm b})$, $\mbox{\Large{$ \varepsilon $}}_{\rm n} (n_{\rm b})$ as a function of baryon density throughout the core \emph{and} the crust, as well as the crust-core transition baryon density $n_{\rm cc}$ and the mesoscopic effective mass of neutrons in the crust $m_{\rm n}^*(n_{\rm b})$.

In order to calculate the crust and core EOSs a model for uniform nuclear matter is required. Nuclear matter models can be characterized by their behavior around nuclear saturation density $n_0 = 0.16$ fm$^{-3}$, the density region from which much of our experimental information is extracted. We can denote the energy per particle of nuclear matter around saturation density by $E(n,\delta)$, where $n$ is the baryon density and $\delta = 1-2x$ the isospin asymmetry, where $x$ is the proton fraction. $x=0.5, \delta = 0$ corresponds to symmetric nuclear matter (SNM), and $x=0, \delta = 1$ to pure neutron matter (PNM). By expanding $E(n, x)$ about $\delta = 0$ we can define the \emph{symmetry energy} $S(n)$,

%===========================================================>
%
%					FIGURE 3
%
%===========================================================>
\begin{figure*}\label{fig:3}
\begin{center}
\includegraphics[width=6cm,height=5cm]{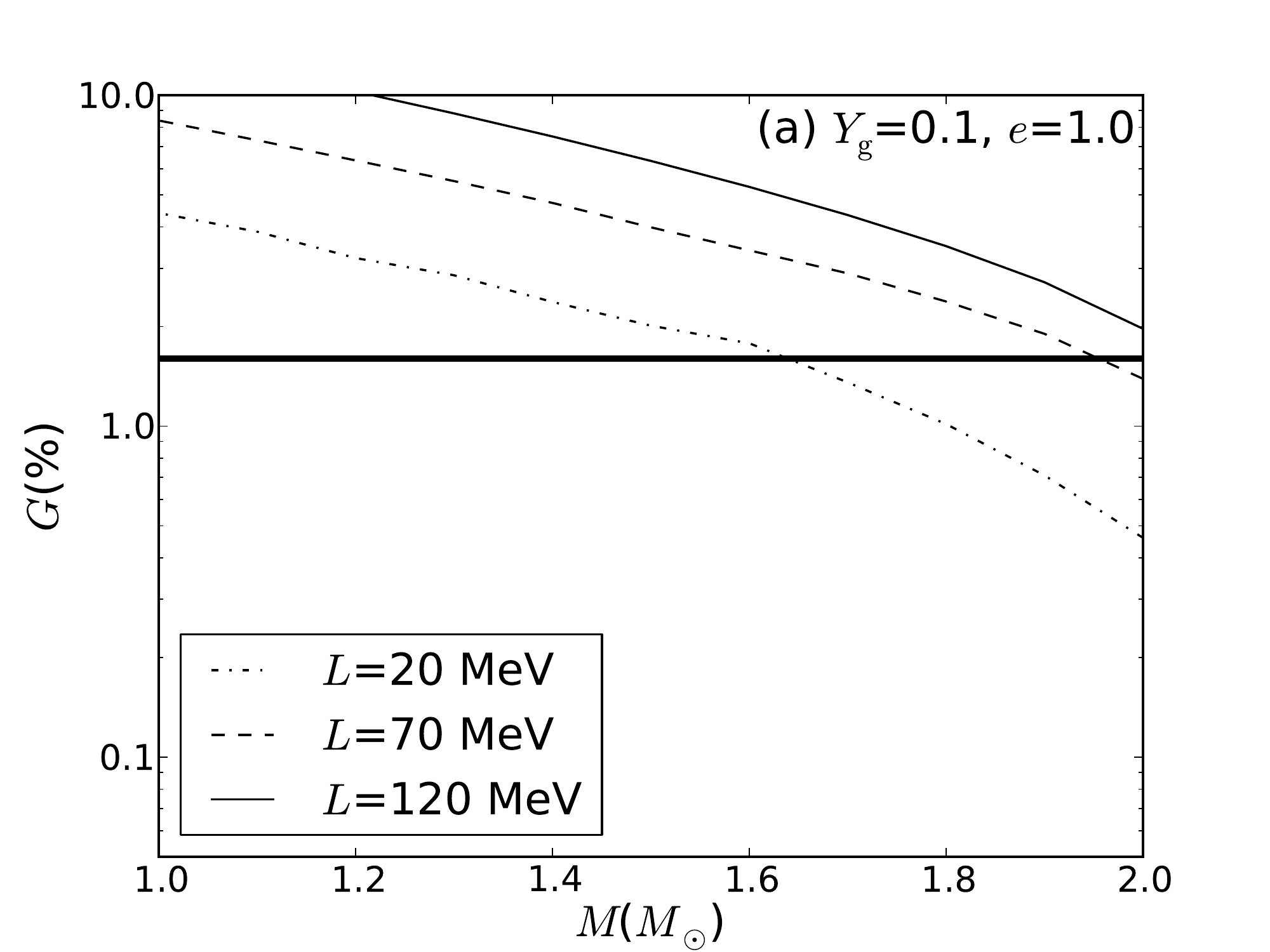}\includegraphics[width=6cm,height=5cm]{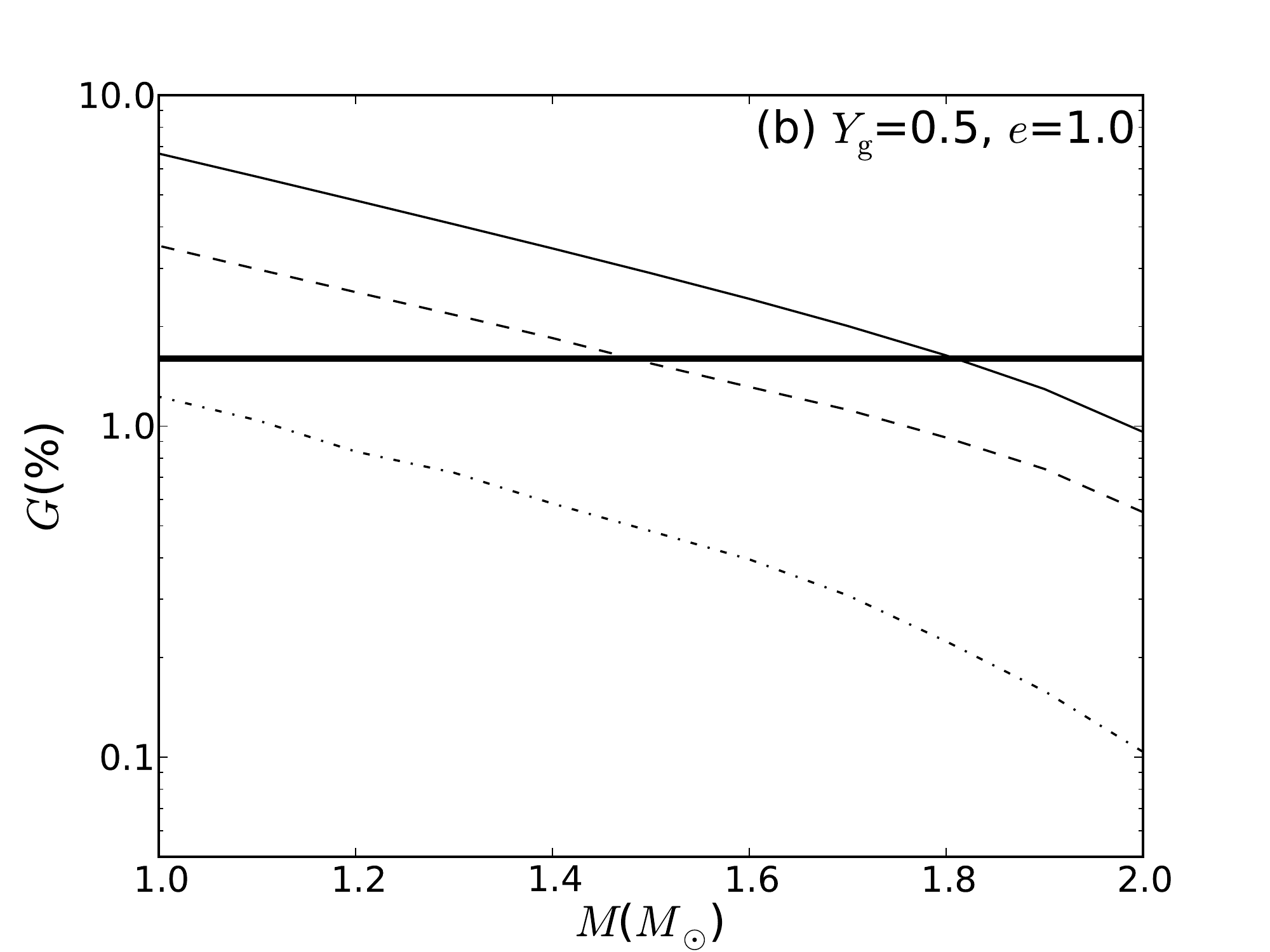}\includegraphics[width=6cm,height=5cm]{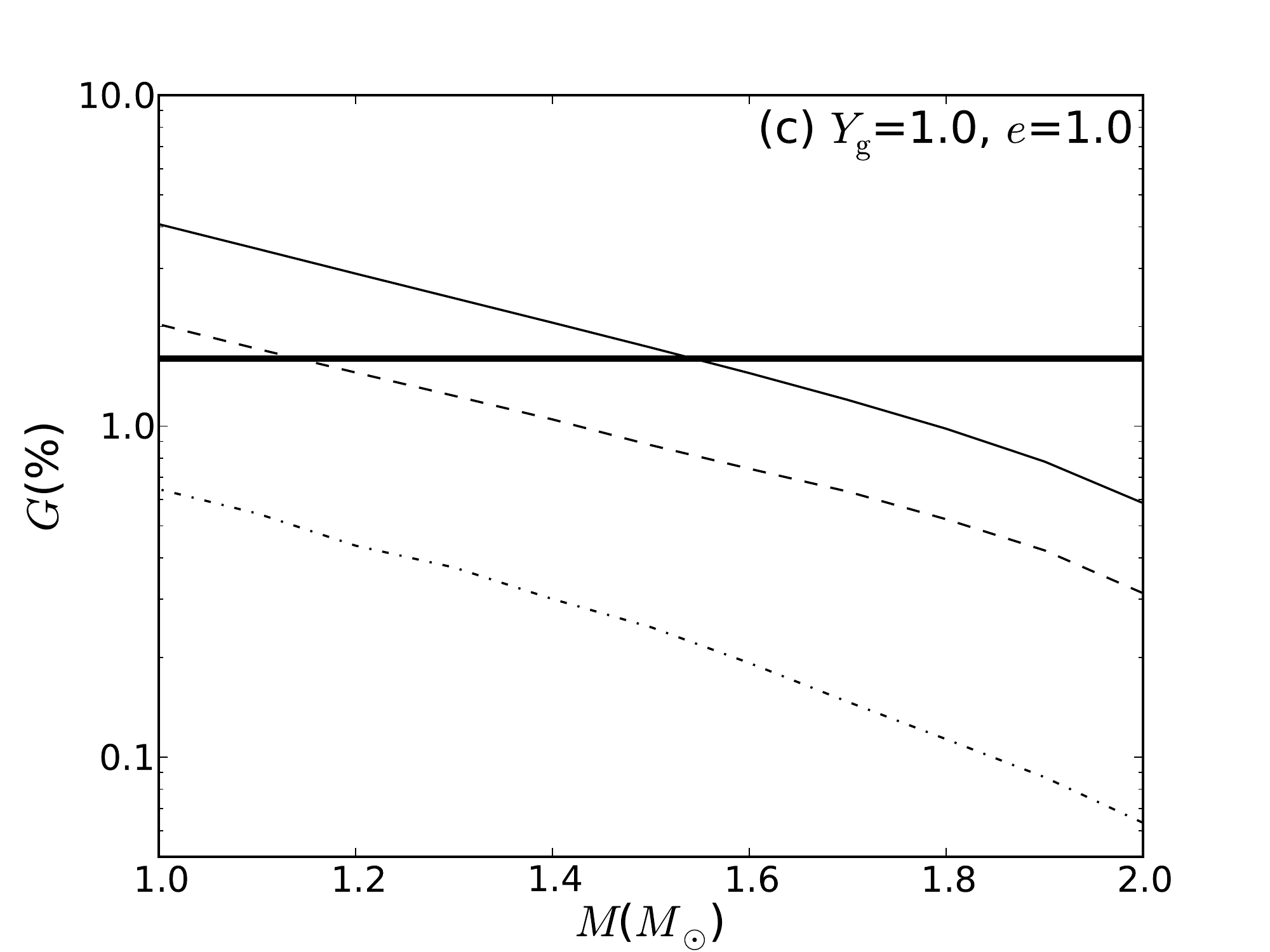}
\includegraphics[width=6cm,height=5cm]{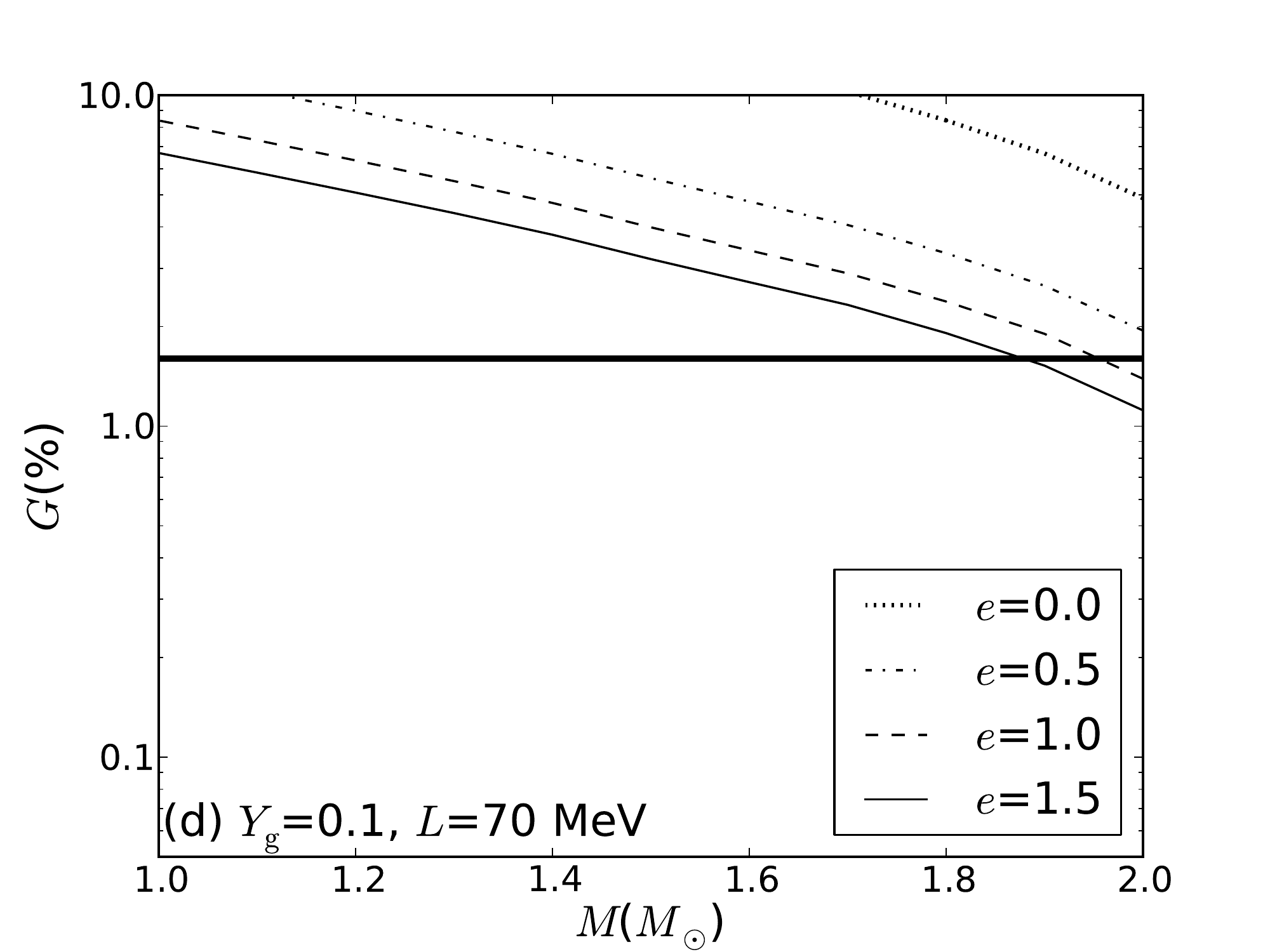}\includegraphics[width=6cm,height=5cm]{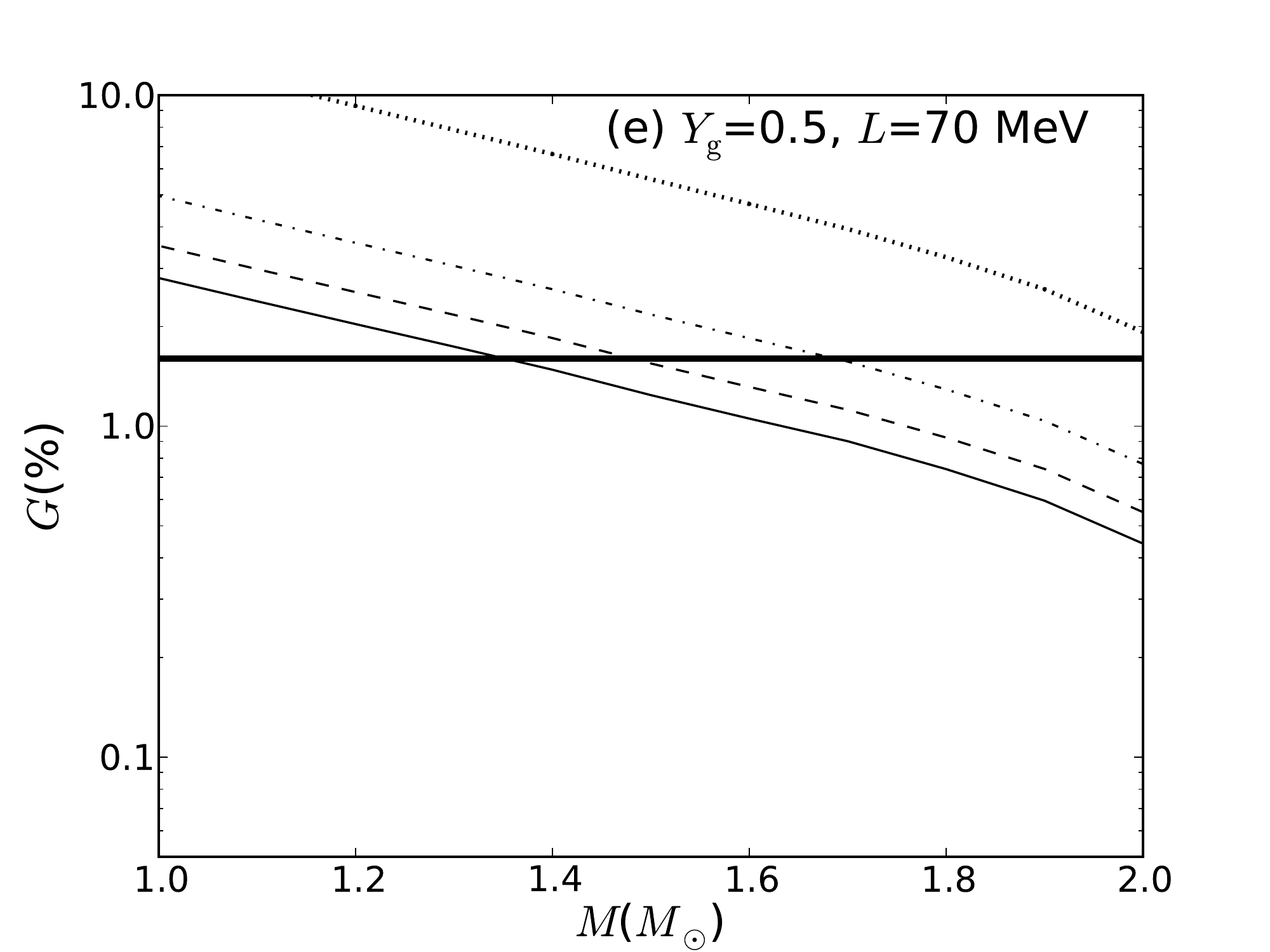}\includegraphics[width=6cm,height=5cm]{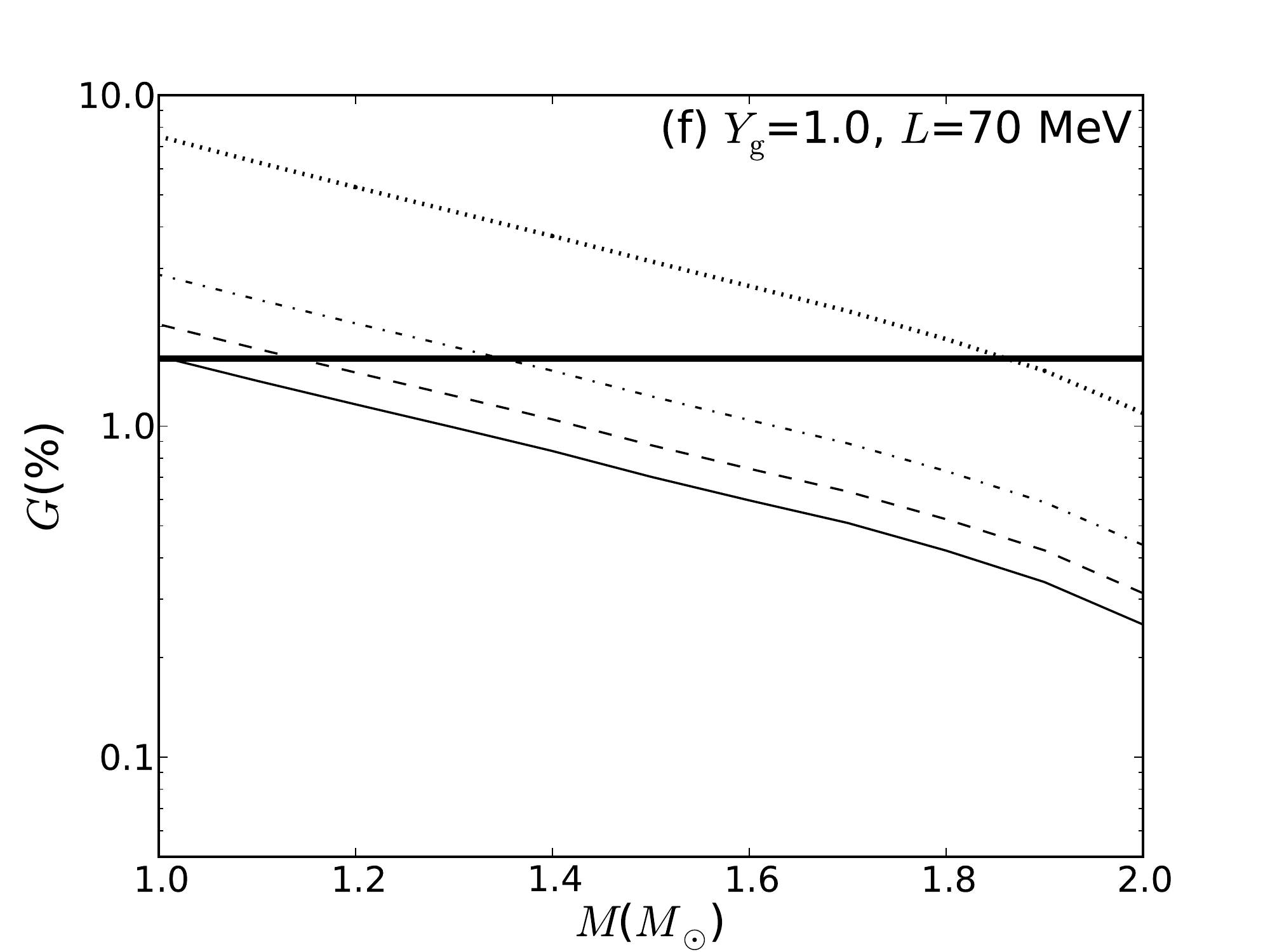}
\caption{The percentage of the moment of inertia of the superfluid neutrons in the inner crust relative to that of the component of the star coupled to the crust at time of glitch, ($G$), versus mass for fractions of core neutrons coupled to the crust $Y_{\rm g}$ = 0.1 (a,d), 0.5 (b,e) and 1.0 (c,f). The top row (a-c) show results $L=20,70,120$ MeV (dash-dot, dashed and solid lines respectively), with full entrainment ($e$=1). The bottom row (d-f) shows results for $L=70$MeV without entrainment $e=0$ (dotted lines), reduced entrainment $e$=0.5 (dash-dot lines), full entrainment $e$=1.0 (dashed lines) and enhanced entrainment $e$=1.5 (solid lines). We show the inferred lower limit to $G$ from the Vela pulsar as the bold horizontal line.}\end{center}
\end{figure*}
%===========================================================>

\be\label{eq:eos1}
E(n,\delta) = E_{\rm 0}(\chi) + S(n)\delta^2 + ...,
\ee
\noindent which encodes the energy cost of increasing the isospin asymmetry of matter. Expanding the symmetry energy about $\chi=0$ where the density parameter $\chi = \frac{n-n_{\rm 0}}{3n_{\rm 0}}$, we obtain
\be\label{eq:eos3}
	S(n) = J + L \chi + \half K_{\rm sym} \chi^{2} + ..., %\frac{J_{sym}}{6}\chi^{3} + \frac{I_{sym}}{24}\chi^4 + ...
\ee

\noindent where $J$, $L$ and $K_{\rm sym}$ are the symmetry energy, its slope and its curvature at saturation density. At present, the energy of SNM around saturation density is well constrained by experiment. Much experimental effort has been focused on determining the symmetry energy $J$ and its density dependence $L$ around nuclear saturation density \citep{Li:2008gp,Tsang:2012se}, and from these results we take as a conservative range $20 < L < 120$ MeV in this work. We will also pay attention, however, to the fact that the congruence of the experimental results \citep{Hebeler2013a,Lattimer2014a} favors a range $30<L<60$ MeV.

We calculate the crust and core EOSs and the transition density consistently using the widely used Skyrme nuclear matter model. As the baseline Skyrme parameterization, we choose the SkIUFSU model used in previous work \citep{Fattoyev:2012ch,Fattoyev:2012uu}, which shares the same saturation symmetric nuclear matter (SNM) properties as the relativistic mean field (RMF) IUFSU model \citep{Fattoyev:2010mx}, has isovector nuclear matter parameters obtained from a fit to state-of-the-art PNM calculations, and describes well the binding energies and charge radii of doubly magic nuclei \citep{Fattoyev:2012ch}.  Two parameters in the Skyrme model are purely isovector - they can be systematically adjusted to vary the symmetry energy $J$ and its density slope $L$ at saturation density while leaving SNM properties unchanged \citep{Chen:2009wv}. The constraint from pure neutron matter at low densities induces correlation between the magnitude and slope of the symmetry energy at saturation density described by $J = 0.167 L + 23.33$ MeV. In this work we will create EOSs by varying $L$ between 20MeV and 120MeV under this constraint. For softer symmetry energies at high densities, the resulting EOS is matched onto two successive polytropic equations of state as described in \citep{Steiner:2010fz,Wen:2011xz} in order to match the constraint on the maximum mass of $M \gtrsim 2$M$_{\odot}$ \citep{Demorest:2010bx,Antoniadis:2013}.

The crust EOS and crust-core transition densities used in this work are obtained from a simple compressible liquid drop model (CLDM) for the crust \citep{Newton:2011dw}. This model gives the composition of the crust (including the free neutron density) and the extent of the so-called `pasta' phases, in which nuclei become deformed into exotic shapes.

To illustrate the relevant correlations between nuclear matter properties and neutron star properties, we plot in Fig.~1 as a function of $L$ the crust-core transition densities $n_{\rm cc}$ and pressures $P_{\rm cc}$ (left plot) and the radii $R$ and inner crust thicknesses $\Delta R_{\rm inner}$ of a $1.4 M_{\odot}$. Although $n_{\rm cc}$ decreases with $L$, the relevant quantity that determines the crust thickness for a star of a given radius is $P_{\rm cc}$, which increases with $L$. A larger stellar radius $R$ also gives a thicker crust, and the right plot shows the well-known positive correlation of $R$ with $L$. These two trends together give the increase of $\Delta R_{\rm inner}$ with $L$ shown in the right plot. Over the range $L=20-120$ MeV, the radius increases by just over 2km and the inner crust thickness triples from approximately 300m to 1000m. Thus a stiffer symmetry energy will give a larger proportion of crust neutrons and we thus expect the capacity of the crust to deliver giant glitches will increase with $L$. 

%===========================================================>
%
%					FIGURE 4
%
%===========================================================>
\begin{figure*}\label{fig:4}
\begin{center}
\includegraphics[width=6cm,height=5cm]{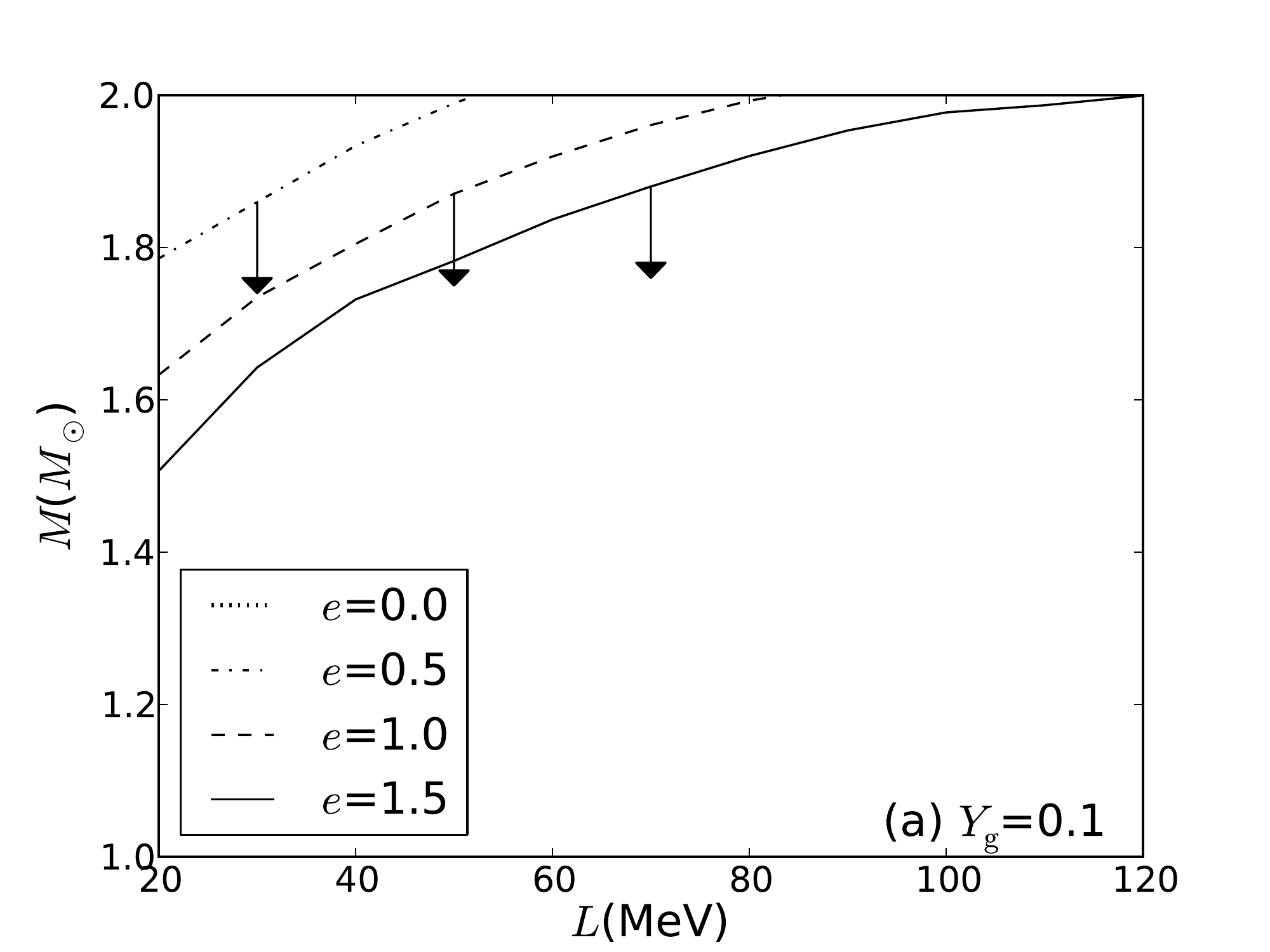}\includegraphics[width=6cm,height=5cm]{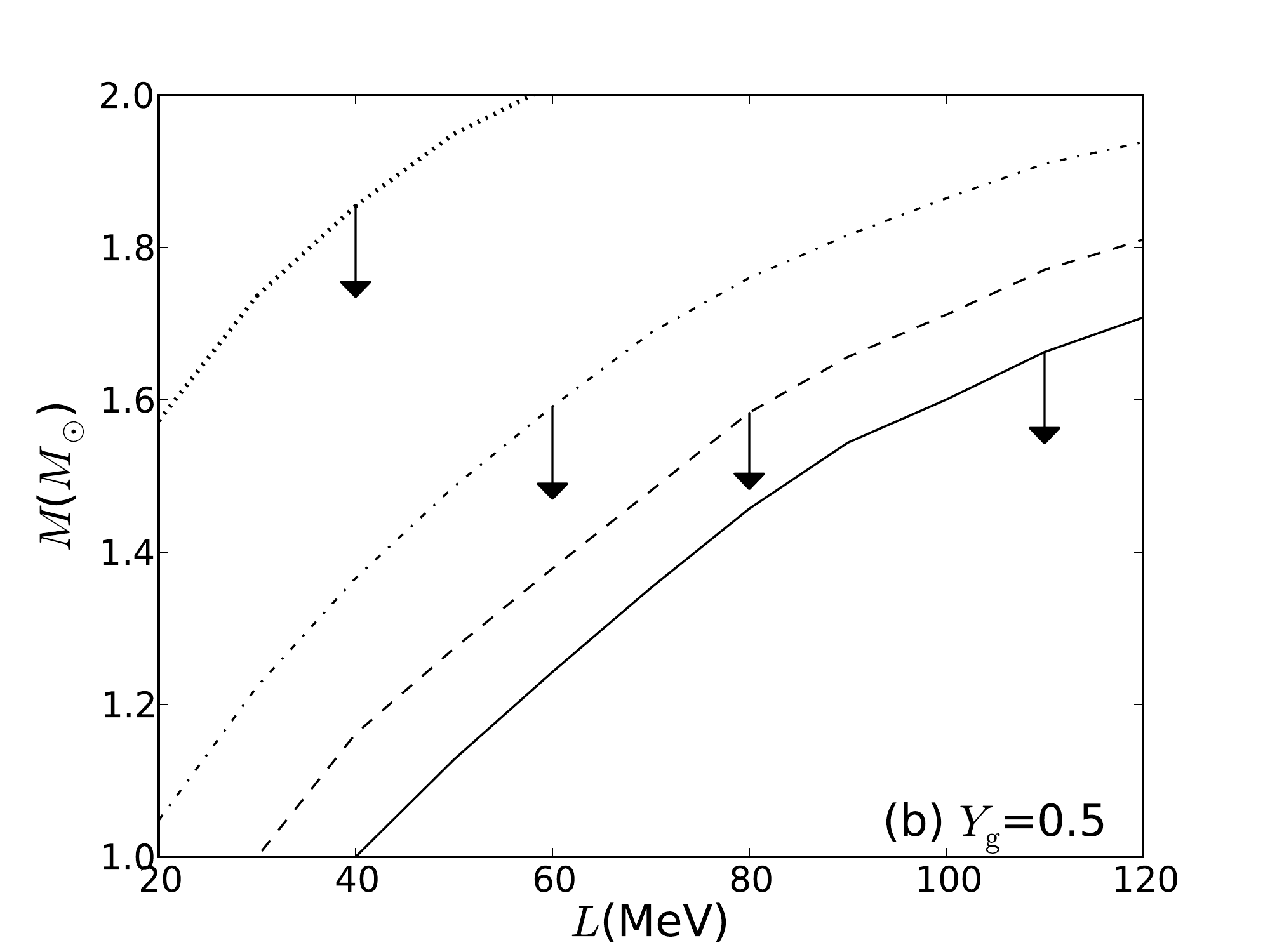}\includegraphics[width=6cm,height=5cm]{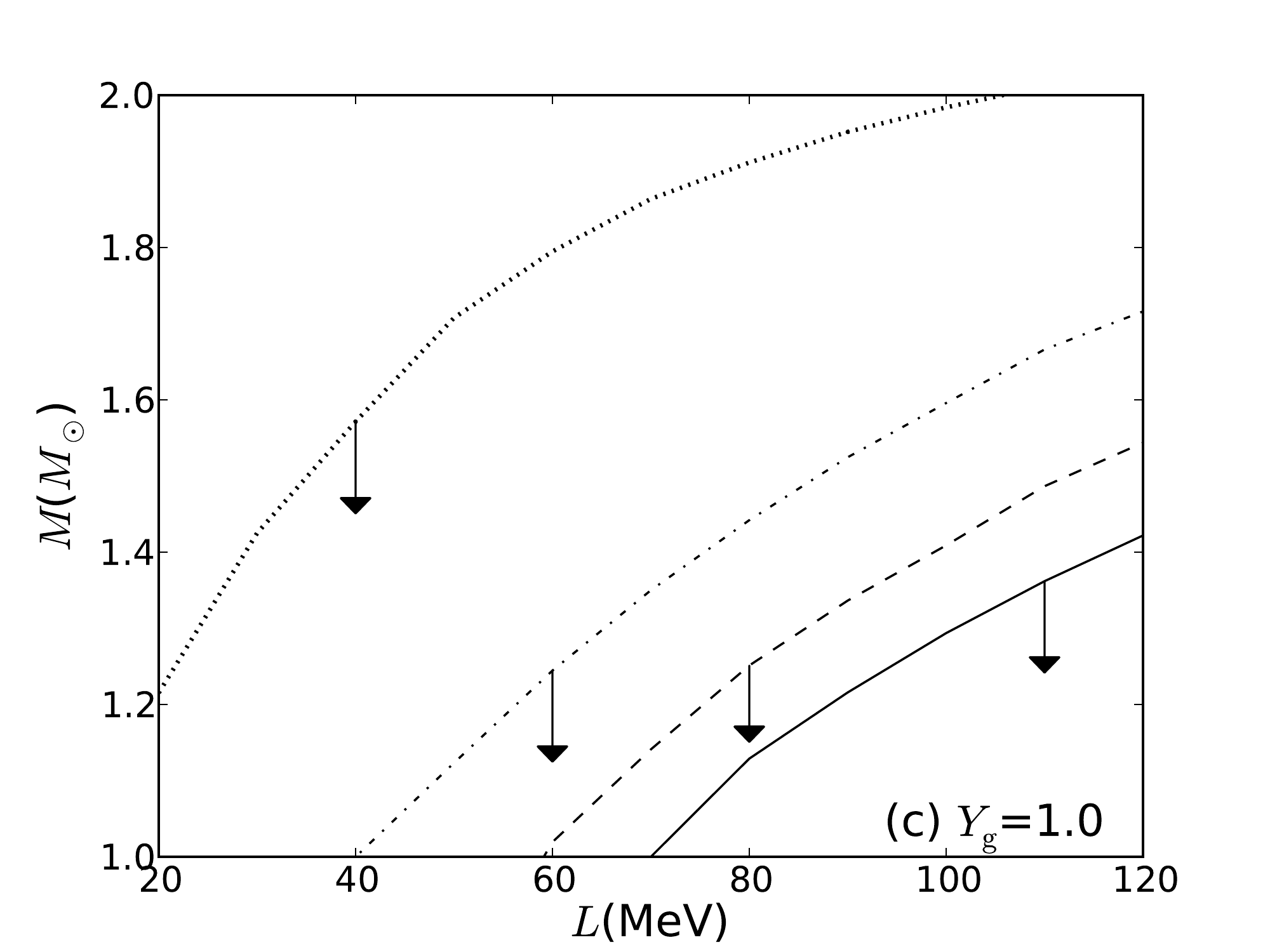}
\caption{Upper limit on the mass for which the Vela constraint $G \geq1.6\%$ is satisfied for a given value of $L$. The region that satisfies $G>1.6\%$ lies below each line as indicated by the arrows. In each plot, results are shown for $e=0$ (dotted line), 0.5 (dash-dotted line), 1.0 (dashed line) and 1.5 (solid line). Results for coupled fractions of core neutrons $Y_{\rm g}$=0.1 (a),  $Y_{\rm g}$=0.5 (b),  $Y_{\rm g}$=1.0 (c) are displayed.}
\end{center}
\end{figure*}
%===========================================================>

Fig.~2 shows, as a function of the radial distance from the center of a $M=1.4 M_{\odot}$ neutron star, the total enthalpy density $h = P + \mbox{\Large{$ \varepsilon $}}$ and that of the neutrons only. In the inner crust, we include the free neutrons \emph{and} those bound in nuclei, and thus our results represent an upper limit to those that would be obtained using free neutrons alone.  We show the enthalpy profiles for $L=25$MeV (a), $L=65$MeV (b), and $L=105$MeV (c). The location of the crust core transition density is indicated, as is the location of the density $0.05$fm$^{-3}$ above that, because we will later on consider extending the angular moment reservoir into the outer core. The previously discussed trends of $R$ and $\Delta R_{\rm inner}$ with $L$ are evident once more. A higher value of $L$ gives a higher value of the symmetry energy at super-saturation densities; therefore the proton fraction in the core $1-Q(r)$ increases with $L$, increasing the moment of inertia of the charged stellar component coupled to the crust at the time of glitch. 

\section{Results and Discussion}

We begin by calculating the ratio of the moment of inertia of the inner crustal neutrons to that of the charged component of the star, $G$=$I_{\rm csf}/I_c$, as a function of mass for a representative range of the model parameters; selected results are displayed in Fig.~3. To recap, the model parameters are: the fraction of core neutrons coupled to the crust at the time of glitch $Y_{\rm g}$, the strength of the crustal neutron entrainment $e$ and the slope of the symmetry energy at saturation density $L$. As mass $M$ increases, so does the compactness of the star $M/R$, while the relative thickness of the crust decreases, giving a smaller contribution to the moment of inertia of the star. Hence the generic trend of $G$ decreasing as mass increases apparent in all plots in Fig.~3.  

In Fig.~3, the inferred lower limit on $G$, 1.6\%, from the Vela pulsar is shown as the horizontal bold line. In Fig.~3 (a-c) we show $G$ as a function of mass $M$ for $Y_{\rm g}$ = 0.1 (a), 0.5 (b) and 1.0 (c), and in each plot we show results for $L=$ 20, 70 and 120 MeV (softest to stiffest symmetry energy). We use entrainment set at the level predicted \citep{Chamel:2012zn}, $e$=1. As $L$ increases (i.e. the symmetry energy at saturation density becomes stiffer), $G$ becomes larger. This relation is the result of the folding together of competing trends. As $L$ increases, the radius of the star and the core proton fraction increases and thus so does the moment of inertia of the charged component of the star $I_{\rm c}$: if the crust thickness were held constant, this would result in a smaller value of $G$. In addition, though, as $L$ increases, the crust thickness increases (Fig.~1b) (as a result of the increasing radius \emph{and} the increasing crust-core transition pressure) thus increasing $I_{\rm csf}$. The overall trend only emerges in a consistent calculation. It is important to note that the emergence of the relation between $G$ and $L$ is very sensitive to the particular relation between the crust-core transition pressure $P_{\rm cc}$ and $L$, which is dependent on the family of EOSs used \citep{Newton:2011dw}. We choose to vary $L$ while maintaining good fits to the results of low-density pure neutron matter calculations. One can choose instead to vary $L$ while fixing the symmetry energy at sub-saturation densities, motivated by fits to nuclear mass models and neutron skin data \citep{Piekarewicz2014a}, and this leads to a different (and potentially non-monotonic) dependence of $P_{\rm cc}$ on $L$, and hence of $G$ on $L$. 

%===========================================================>
%
%					FIGURE 5
%
%===========================================================>
\begin{figure*}\label{fig:5}
\begin{center}
\includegraphics[width=6cm,height=5cm]{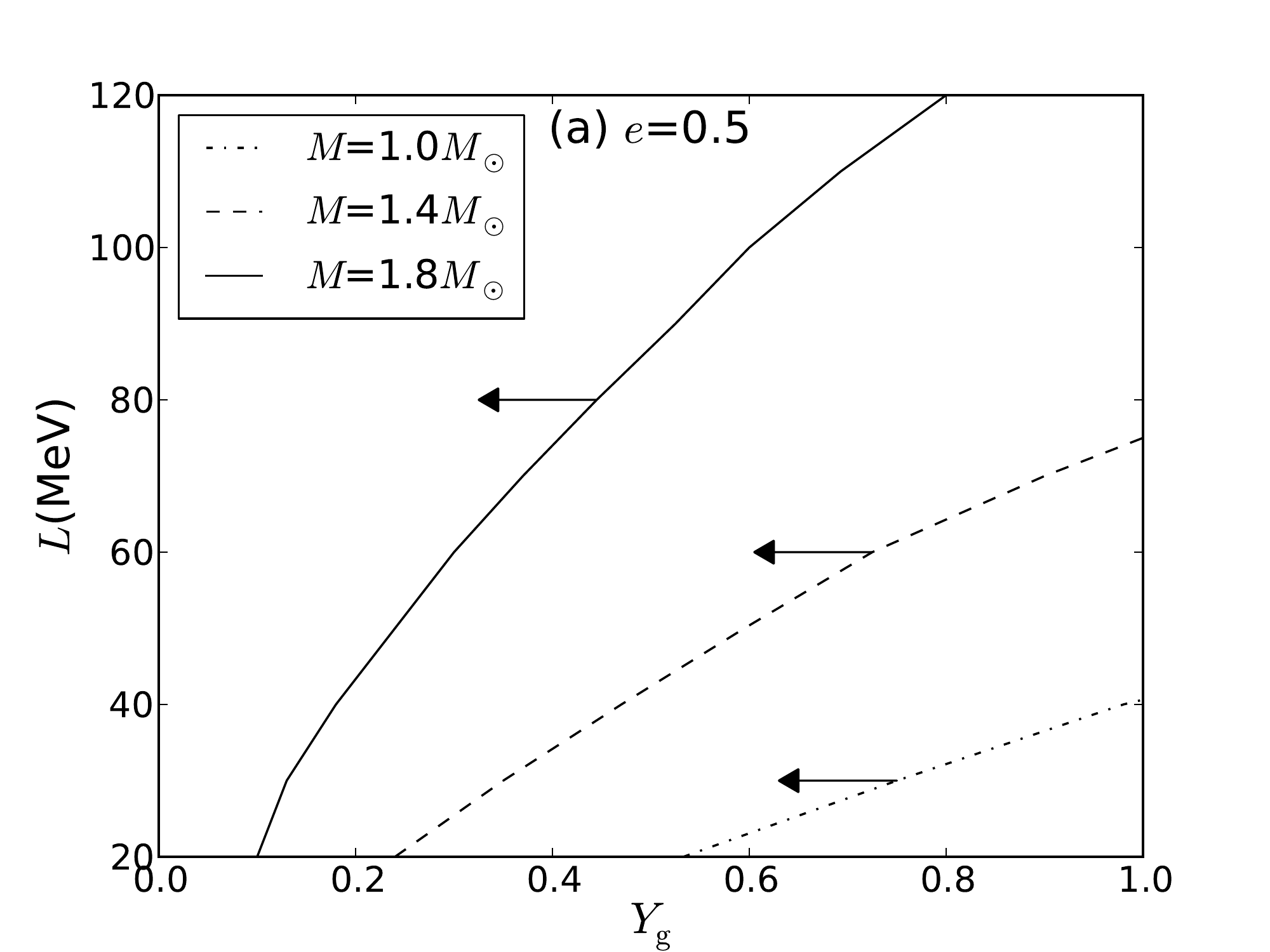}\includegraphics[width=6cm,height=5cm]{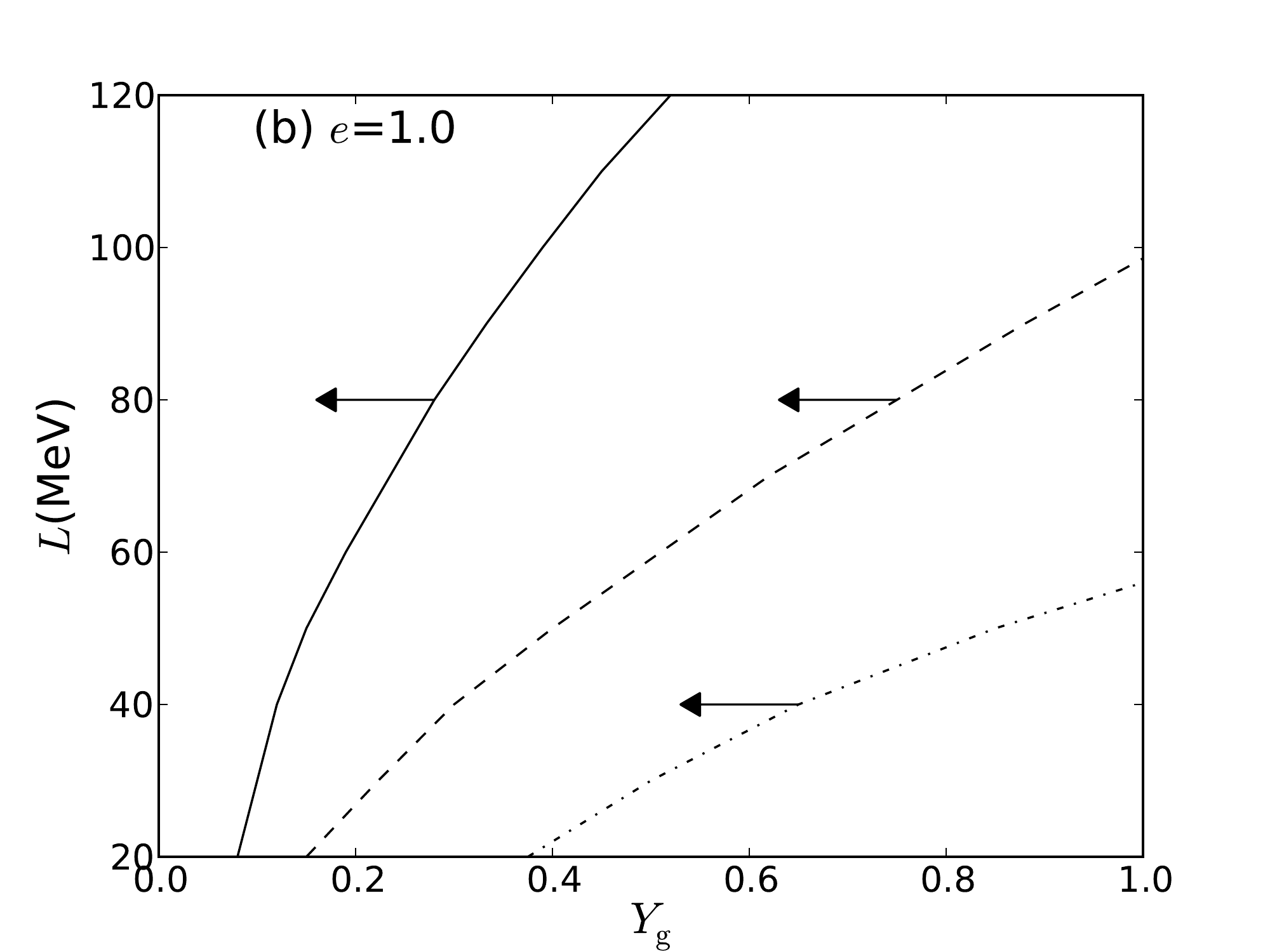}\includegraphics[width=6cm,height=5cm]{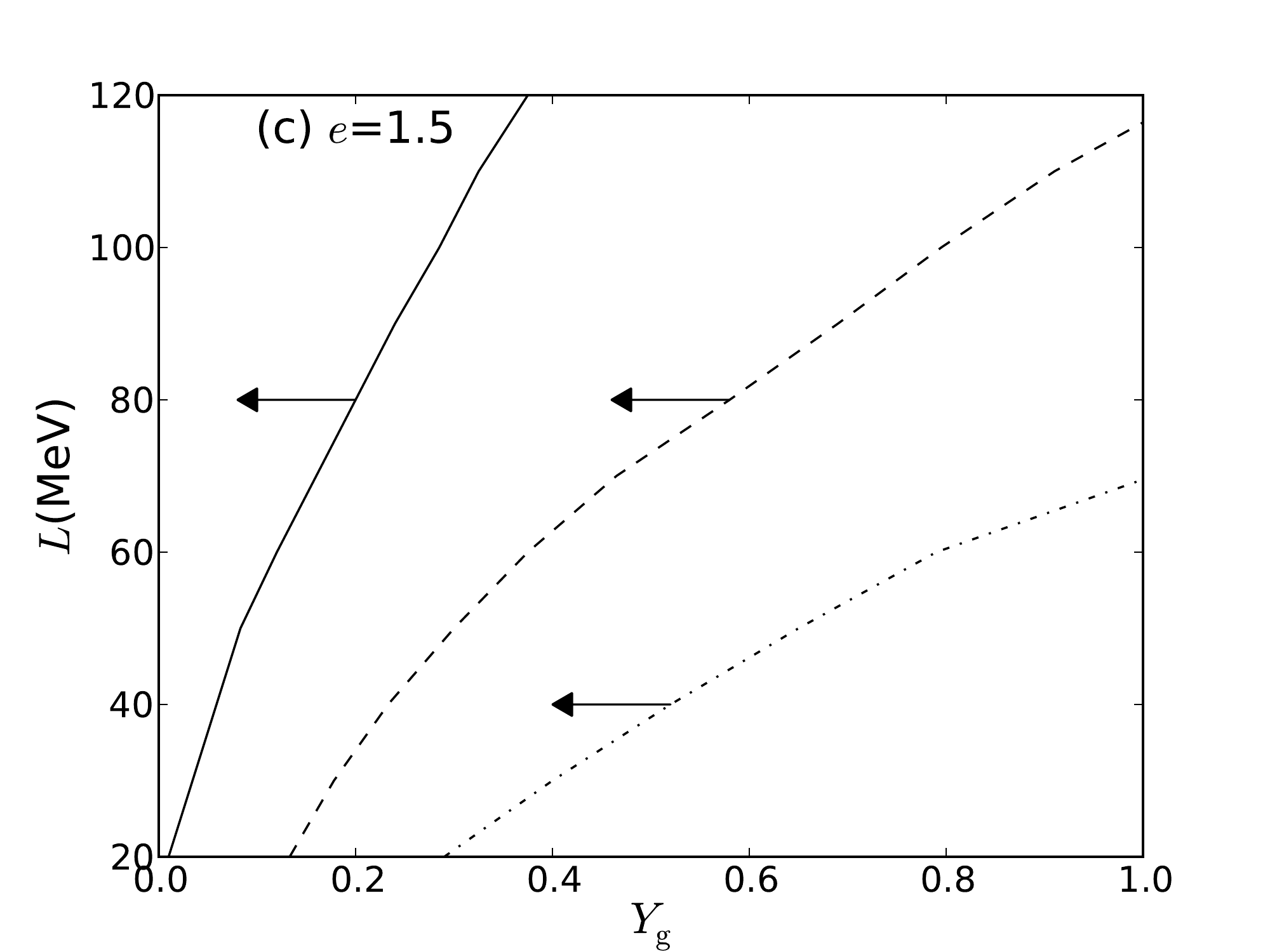}
\caption{Fraction of the core neutrons coupled to the crust at the time of glitch below which $G\geq1.6\%$ is satisfied for a given value of $L$. Results are shown for reduced entrainment $e$=0.5 (a), full entrainment $e$=1.0 (b) and enhanced entrainment $e=1.5$(c) for a 1.0$M_{\odot}$ star (dash-dotted lines), a 1.4$M_{\odot}$ star (dashed lines) and for a 1.8$M_{\odot}$ star (solid lines). Regions of parameter space consistent with $G\geq1.6\%$ lie to the \emph{left} of a given curve as indicated by the arrows.}
\end{center}
\end{figure*}
%===========================================================>

Increasing $Y_{\rm g}$ for a fixed mass star increases the fraction of the core superfluid that couples to the crust at the time of glitch and thus $I_{\rm c}$, therefore decreasing $G=I_{\rm csf}/I_{\rm c}$. If the entire core is taken to be coupled to the crust, as it is in the analyses of \citep{Chamel:2012ae,Piekarewicz2014a,Andersson:2012iu}, then we see from Fig.~3c that we require stiff EOSs $L\gtrsim$100 MeV to obtain $G>$1.6\% for a 1.4$M_{\odot}$ star, in line with the analyses of \citep{Chamel:2012ae,Piekarewicz2014a}. These are stiffer EOSs than are currently inferred from most experimental data. With full entrainment, we have to lower the fraction of core neutrons coupled to the crust to $Y_{\rm g}\sim$0.5 in order to satisfy $G>$1.6\% for a 1.4$M_{\odot}$ star for EOSs in the range $L\lesssim$70 MeV, more in line with experimental constraints. A very low fraction of core neutrons coupled to the crust $Y_{\rm g}$ allows the whole range of EOSs to comfortable satisfy $G>$1.6\% for a 1.4$M_{\odot}$ star.

In Fig.~3 (d-f) we consider a range of possible strengths of entrainment relative to that calculated: no entrainment $e$=0, reduced entrainment $e$=0.5, full entrainment $e$=1.0 and enhanced entrainment $e=1.5$. Results are displayed for $L$=70 MeV over the same range of $Y_{\rm g}$. Note that since the moment of inertia of the crustal neutrons which participate in the glitch depends inversely on the strength of entrainment, increasing that strength in equal increments has successively smaller effect on the predicted value of $G$. Increasing entrainment by 50\% compared to \citep{Chamel:2012zn} gives a relatively small reduction in $G$; at $Y_{\rm g}$=0.5, our $L$=70 MeV EOS is still able to just satisfy $G>$1.6\% for a 1.4$M_{\odot}$ star. Equally, reducing entrainment by 50\% still requires relatively stiff EOSs to satisfy $G>$1.6\%; for a fully coupled core $Y_{\rm g}$=1.0, $L$=70 MeV only marginally satisfies the constraint for $e$=0.5.

For a given set of model parameters, we can identify a mass $M$ below which the Vela constraint on the moment of inertia ratio, $G\geq1.6\%$, is satisfied. These upper mass limits are plotted as a function of $L$ in Figs.~4a-c. Results are shown $Y_{\rm g}$ = 0.1 (a), 0.5 (b) and 1.0 (c). In each plot, the lines represent the upper limit on the mass as a function of $L$ for entrainment strengths of $e=$0, 0.5, 1.0 and 1.5. Masses which satisfy $G\geq1.6\%$ lie \emph{below} the curve, as indicated by the arrows. 

When entrainment of crust neutrons is neglected $e=0$, $G\geq1.6\%$ is satisfied for all values of the slope of the symmetry energy $L$ for masses below $1.2M_{\odot}$ for $Y_{\rm g}=1.0$, below $1.6M_{\odot}$ for $Y_{\rm g}=0.5$ and for all masses considered up to $2.0M_{\odot}$ for $Y_{\rm g}=0.1$. For low coupled core neutron fractions down to $Y_{\rm g}=0.1$, the Vela constraint is satisfied for masses less than $1.5M_{\odot}$ for all EOSs even with enhanced entrainment $e$=1.5. For a fully coupled core, even reduced entrainment restricts significantly the range of EOSs that can satisfy the Vela constraint to $L\geq$70 MeV for canonical neutron star masses $1.4M_{\odot}$.

For each EOS (that is, for each value of $L$) one can find an upper limit on $Y_{\rm g}$ for which $G\geq1.6\%$ is satisfied below a given mass of neutron star. Fig.~5a-c plots those values of $L$ vs $Y_{\rm g}$; only values of $Y_{\rm g}$ \emph{to the left} of the curves satisfy the observational constraint on $G$ as indicated by the arrows. We show results for reduced entrainment $e$=0.5 (a), full entrainment $e$=1.0 (b) and enhanced entrainment $e$=1.5 (c). Each plot shows results for masses of 1.0, 1.4 and 1.8$M_{\odot}$. As one can see from Fig.~5c, we are able to satisfy the Vela constraint for any EOS for stars of mass $1.4M_{\odot}$ and below if the coupled core neutron fraction is below $Y_{\rm g}=0.13$ even with enhanced entrainment. For full entrainment, to be consistent with \emph{any} EOS in the range $L\approx$ 30-60 MeV from the concordance of a number of experimental results, then for full entrainment $e$=1.0 we require $Y_{\rm g}\lesssim$0.2 for a $1.4M_{\odot}$ star. For a high mass $1.8M_{\odot}$ star, this is reduced to  $Y_{\rm g}\lesssim$0.1. If we only require consistency with an $L=60$MeV EOS, we require $Y_{\rm g}\lesssim$0.5 for a $1.4M_{\odot}$ star and $Y_{\rm g}\lesssim$0.2 for a $1.8M_{\odot}$ star.

There is a model dependence in the calculation of the crust-core transition density and pressure from both the EOS and the method of calculation. Indeed, a different method of calculation and a different set of EOSs has been shown to be able to satisfy the Vela constraint for values of $L$ in line with experimentally inferred values \citep{Piekarewicz2014a}. In addition, it is possible that core neutrons can also be involved in driving the glitch. We therefore consider increasing the density range over which we calculate the moment of inertia of neutrons responsible for the glitch by an amount $n^+$. Taking the canonical model of a fully coupled core $Y_{\rm g}$=1.0 and a $1.4M_{\odot}$ star, we calculate the size of the density increase $n^+$ we must apply in order to satisfy the Vela constraint for each value of $L$. The results are displayed in Fig.~6 for reduced entrainment $e$=0.5, full entrainment $e$=1.0 and enhanced entrainment $e$=1.5. One important point can be taken from these results. In order to satisfy the Vela constraint in the canonical model for $L\approx$ 30-60 MeV, we require an increase in density of $\sim$0.01fm$^{-3}$ above our crust-core transition densities \emph{even with enhanced entrainment} . This is within the reasonable uncertainties in the crust-core transition density \citep{Newton:2011dw, Piekarewicz2014a}, indicating that even with $Y_{\rm g}$=1.0 we cannot rule out the purely crustal origin of glitches. Alternatively, only a small amount of outer core neutrons are required to contribute to the angular momentum transfer to the crust (perhaps through pinning on type II superconducting flux tubes) in order to satisfy the Vela constraint. 

%===========================================================>
%
%					FIGURE 6
%
%===========================================================>
\begin{figure}\label{fig:6}
\begin{center}
\includegraphics[width=8cm,height=6cm]{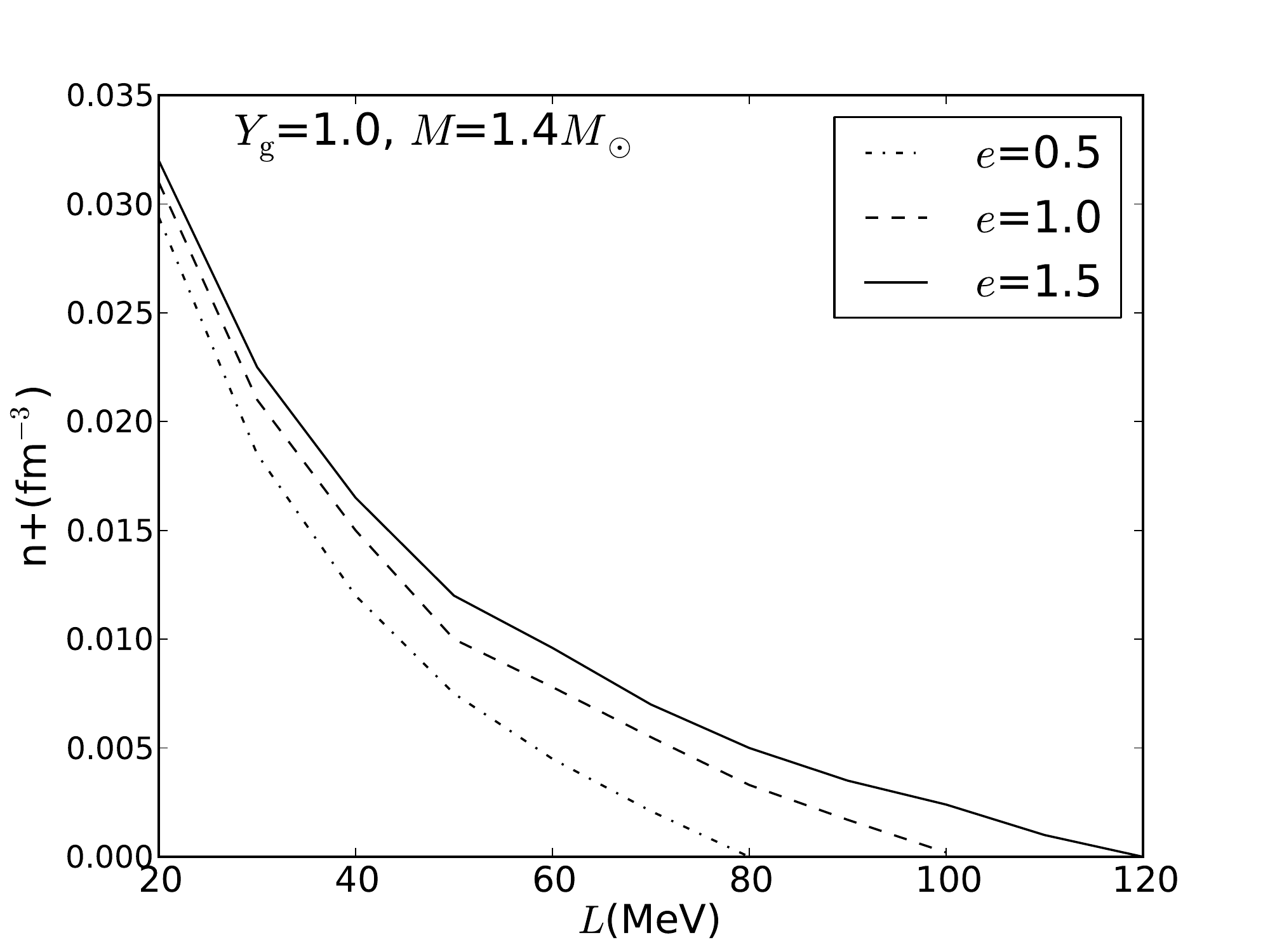}
\caption{The increase in density $n^+$ above the crust-core transition to which we must extend the region of neutrons which act as the angular momentum reservoir for the glitch in order to satisfy the Vela constraint. Results are shown for a fully coupled core $Y_{\rm g}$=1.0, a 1.4$M_{\odot}$ star and for reduced entrainment $e$=0.5 (dash-dotted line), full entrainment $e=1.0$ (dashed line) and enhanced entrainment $e$=1.5 (solid line).}
\end{center}
\end{figure}
%===========================================================>

\section{Conclusions}

Using neutron star crust and core EOSs and compositions calculated using the same Skyrme nuclear matter model we have explored whether crust neutrons provide a sufficient angular moment reservoir to reproduce the observed glitch activity of the Vela pulsar allowing for the reduction of free neutrons by crustal entrainment. We go beyond similar studies by allowing for only a fraction of core neutrons to be coupled to the crust on the timescale of the glitch. The Skyrme model contains two parameters that allow for systematic variation of the symmetry energy magnitude and slope at saturation density without changing symmetric nuclear matter properties, which we exploit to examine predictions spanning a conservative range of uncertainty in the density-slope of the symmetry energy $20<L<120$ MeV, while simultaneously being constrained to fit the results of state-of-the-art pure neutron matter calculations at low densities. Other than identifying the crust neutrons as the angular momentum reservoir, our results are independent of the details of the glitch mechanism. We compare our results to the average glitch activity of the Vela pulsar, which results in the most stringent constraint, $G\geq1.6\%$.

We explore variations in the following model parameters: the slope of the symmetry energy $L$, the fraction of core neutrons coupled to the crust at the time of glitch $Y_{\rm g}$, the strength of the crustal neutron entrainment $e$ relative to the results of Bragg scattering of neutrons off crustal nuclei \citep{Chamel:2012zn}. We take as the baseline model $Y_{\rm g}$ = 1.0,  $e$=1 and a 1.4$M_{\odot}$ star.

For our baseline model, we find that we require stiff EOSs $L>100$MeV to generate glitch activities at the level observed in the Vela pulsar for a 1.4$M_{\odot}$ star. This is in agreement with the results of  \citep{Chamel:2012ae,Andersson:2012iu}, although we have now quantified the required stiffness in terms of $L$. Comparison with the tentative range $L\approx$30-60 MeV from the congruence of several experimental probes \citep{Hebeler2013a,Lattimer2014a} suggests that if the whole core is coupled to the crust, the crust neutrons are not sufficient to drive the glitch.

Two other studies have conducted systematic explorations of equation-of-state parameter space in order to determine compatibility with the Vela glitch activity \citep{Piekarewicz2014a,Steiner2015a}. Both studies find \emph{some} models that are compatible even with entrainment in full effect. In both cases, such models occupy a small and specific region of parameter space, but it is nevertheless necessary to outline the difference between these models and those used in this study.

\cite{Piekarewicz2014a} arrived at the conclusion that the crust neutrons can drive large enough glitches for moderate values of $L\approx$70 MeV. The core EOSs used in that work are derived using families of parameterizations of relativistic-mean field models; in our work, we use a parameterization of the non-relativistic Skyrme model that is fit at first to a particular parameterization of one such RMF model, and therefore reproduces its predicted observables. Both RMF and Skyrme models contain two purely isovector parameters - that is, two parameters whose adjustment affects only predictions of observables sensitive to neutron-proton asymmetries. Starting from our baseline parameterization, SkIUFSU we re-fit both purely isovector parameters in order to match the results of low-density pure neutron matter (PNM) calculations  \citep{Gezerlis:2009iw, Hebeler:2009iv, Gandolfi:2011xu} (which we emphasize are expected to be quite robust.). Subseqeuntly, we vary $L$ by adjusting one of the isovector parameters, while adjusting the second to maintain our fit to the PNM calculations.

The study of \cite{Piekarewicz2014a} starts from a number of baseline RMF parameterizations, all the results of fits to slightly different sets of experimental data as well as, in the case of the one model that satisfies the Vela glitch activity, a hypothetical neutron skin measurement on $^{208}$Pb. Subsequently, only one of the two isovector parameters is adjusted (an adjustment which is tracked by the resulting change in the neutron skin thickness of $^{208}$Pb) in order to find the value that gives the highest value of the crust-core transition pressure $P_{\rm t}$ and hence the best chance of reproducing the Vela glitch activity. The second isovector parameter remains fixed during this. There is no comparison with the low density PNM EoS, but such a procedure will result only in accidental agreement with the results of PNM calculations.

The matching to PNM matter becomes particularly important when we consider the second main difference between the work of \cite{Piekarewicz2014a} and that presented here. We use a model of the inner crust composition and EOS, and crust-core transition pressure, that is fully consistent with the core EOS in that it is derived from the same underlying Skyrme model. The key component of the inner crust EOS in these studies is the low-density neutron fluid in the interstices of the crustal lattice; it is therefore important to use our best knowledge of the low-density PNM EOS when constructing such models, and ensure it is free of artifacts (such as, for example, bound neutron matter). The work of \cite{Piekarewicz2014a}, while calculating the crust-core transition pressure consistently, uses a polytropic EOS for the inner crust, which is matched onto the core EOS; such an EOS will necessarily be inconsistent from a nuclear physics point of view with the core EOS, and any low-density artifacts that may arise in their EOSs as a result of ignoring the low-density EOS in PNM matter will remain hidden. Without a direct comparison with their models, we cannot comment on what difference this will make on the predictions of the crustal moment of inertia.

The second study to conduct a systematic exploration of EOS parameter space \citep{Steiner2015a} does use the results of low-density PNM matter calculations to parameterize the low-density EOSs used; they also use the same crustal EOSs as used here. Nevertheless, they do find a small portion of parameter space consistent with Vela glitch activity at moderate values of $L=30-60$MeV. The reason lies in the high density EOS, which are polytropes in that work. The models which are compatible with Vela's glitch activity have particularly stiff EOSs in the higher density regions of the core, giving rise to much larger radii and hence crustal thicknesses and moments of inertia. Such stiff EOSs at high density predict high radii for 1.4$M_{\odot}$ neutron stars, in excess of 14km, and can therefore be tested by future radius measurements. In our work, we use the same EOS at high densities and low densities, and only append a polytrope at the highest densities when we need to stiffen the EOS sufficiently to reproduce a 2$M_{\odot}$ neutron star. Therefore, the higher and lower density regions are more strongly correlated, and we lack much freedom to vary the high density EOS independently of the low density component. Our high density EOS is nevertheless based on a microphysical model, which is lacking from a simple polytropic description. More work needs to be done on exploring the high-density EOS parameter space on a microscopic footing, given is current uncertainty both theoretically and experimentally \citep{Xiao:2009zza,Russotto:2011hq}, even if one restricts the composition to purely nucleonic matter. 

Even taking into account pure neutron matter calculations, there is sufficient uncertainty in the remaining parameters of the nuclear model underlying the EOS to cause uncertainties of $\sim$10\% in the crust-core transition density and pressure \citep{Newton:2011dw}. To allow for this uncertainty, we examine how much the crust-core transition density would have to be increased in order to match the observations. In the range $L\approx$30-60 MeV we find an increase of $n^+\approx$0.01fm$^{-3}$ is required, which is of order the EOS model uncertainties. Thus we must conclude that for the baseline model, it is still marginally possible to match the Vela glitch activity using crust neutrons with EOSs that are also consistent with nuclear experiment \emph{and} the results of low density pure neutron matter calculations. One can also interpret the increase in density $n^+$ as allowing for the participation of core neutrons in the glitch mechanism, in which, given the set of EOS parameters used here, we set a lower limit $n+\approx$0.01fm$^{-3}$ on the density above the crust-core transition that this participation must occur (although this assumes that the participating core neutrons must be contiguous with the crust neutrons, which need not be the case).

We explore the effect of reducing and increasing entrainment for 50\%. For reduced entrainment, we require $L>$75 MeV for $Y_{\rm g}$ = 1.0,  $M$=1.4$M_{\odot}$. For enhanced entrainment, we require $L>$115 MeV. We have noted that increasing entrainment by 50\% (which is equivalent to increasing the neutron effective masses in the crust by 50\%) does not greatly increase the problem of matching the Vela results; an increase in density $n^+\approx$0.01fm$^{-3}$ would still provide enough neutrons to allow for such agreement.

We also consider lower and higher mass stars. If Vela is a high mass neutron star  $M$=1.8$M_{\odot}$, then we cannot match the observed glitch activity for any EOSs considered for $Y_{\rm g}\gtrsim$0.5 for full strength entrainment $e=1.0$.

Allowing for only a fraction of the core neutrons to be coupled to the crust at the time of the glitch, we find that for consistency with experiment $L<$60 MeV and for a 1.4$M_{\odot}$ star, we require $Y_{\rm g}<$0.5 for $e$=1.0,  $Y_{\rm g}<$0.7 for $e=0.5$ and $Y_{\rm g}<$0.4 for $e=1.5$. For a 1.8$M_{\odot}$, we would require $Y_{\rm g}<$0.2 for $e$=1.0,  $Y_{\rm g}<$0.3 for $e=0.5$ and $Y_{\rm g}<$0.1 for $e=1.5$. 

Considering the crust-core coupling to be mediated by core mutual friction, $Y_{\rm g}$ can be related to the strength of mutual friction in the core via the crust-core coupling timescale \citep{Haskell:2011xe,Haskell2014aa} and therefore, in principle, calculated consistently with the EoS, thereby removing it as a free model parameter. This would be a possible next step in a microphysically-consistent analysis of glitch models, to be reported in an upcoming paper. Note that the strength of mutual friction and hence $Y_{\rm g}$ will vary throughout the core as the composition and overall density changes.

To summarize, for a 1.4$M_{\odot}$ star and a fully coupled core, we require very stiff EOSs to match glitch observations at odds with current experiment. These results are consistent with \citep{Chamel:2012ae,Andersson:2012iu} within the set of EOSs used here constrained by pure neutron matter calculations. Allowing for uncertainties in the crust-core transition pressure of order 10\% would be sufficient to remove this problem; alternatively, only a small amount of core neutrons need to be involved to alleviate the problem. If less than $\approx$40\% of the core is coupled to the crust at the time of glitch, 
crust-originated glitches are possible for EOSs consistent with experimental data and pure neutron matter calculations, even if the strength of crustal entrainment is increased to 50\% above its predicted values.

\section*{ACKNOWLEDGEMENTS}
We thank Brynmor Haskell and Pierre Pizzochero for invaluable discussions. We thank the referee for help improving the manuscript. This work is supported in part by the National Aeronautics and Space Administration under grant NNX11AC41G issued through the Science Mission Directorate and the National Science Foundation under grants PHY-0757839 and PHY-1068022.

\bibliographystyle{mn2e}

\end{document}